\documentclass[8pt]{article}

\usepackage[table]{xcolor}

\usepackage{graphicx,pslatex,float,color,array,amssymb,amsmath,euscript}
\usepackage{pxfonts}
\usepackage{ifthen}
\usepackage{longtable}
\usepackage{eso-pic}
\usepackage{sidecap}
\usepackage{xspace}
\usepackage[paperwidth=21.0cm,paperheight=29.7cm,textwidth=16cm,textheight=23cm,top=3cm,left=2.5cm]{geometry}

\definecolor{navy}{rgb}{0,0,0.5}
\usepackage[pdftex,
           hyperindex=true,
           colorlinks=true,
           linkcolor=navy,
           anchorcolor=magenta,
           citecolor=navy,
           urlcolor=navy,
           unicode,
           implicit=true]{hyperref}

\newcommand{\be}{\begin{equation}}
\newcommand{\ee}{\end{equation}}
\newcommand{\bes}{\begin{equation*}}
\newcommand{\ees}{\end{equation*}}

\newcommand{\erf}{\mathrm{erf}}
\newcommand{\erfc}{\mathrm{erfc}}

\newcommand{\ddp}[2]{\frac{\partial #1}{\partial #2}}
\renewcommand{\Re}{\mathrm{Re}}
\renewcommand{\Im}{\mathrm{Im}}

\title{The role of the roughness spectral breadth\\in elastic contact of rough surfaces}

\author{Vladislav A. Yastrebov$^a$\footnote{Corresponding author $\langle$\texttt{vladislav.yastrebov@mines-paristech.fr}$\rangle$}, Guillaume Anciaux$^b$, Jean-Fran\c cois Molinari$^b$}

\date{\small{\it$^a$MINES ParisTech, PSL Research University, Centre des Mat\'eriaux}\\{\it CNRS UMR 7633, BP 87, F 91003 Evry, France}\\
\small{\it
$^b$Computational Solid
   Mechanics Laboratory (LSMS, IIC-ENAC, IMX-STI),}\\{\it Ecole Polytechnique
   F\'ed\'erale de Lausanne (EPFL), B\^at. GC, Station 18, CH 1015
   Lausanne, Switzerland}}

\begin{document}

\maketitle

\begin{flushleft}
 \Large{\bf Abstract.}\normalsize
\end{flushleft}

\noindent 
We study frictionless and non-adhesive contact between elastic half-spaces with self-affine surfaces.
Using a recently suggested corrective technique, we ensure an unprecedented accuracy in computation of the true contact area evolution under increasing pressure.
This accuracy enables us to draw conclusions on the role of the surface's spectrum breadth (Nayak parameter) in the contact area evolution.
We show that for a given normalized pressure, the contact area decreases logarithmically with the Nayak parameter.
By linking the Nayak parameter with the Hurst exponent (or fractal dimension), we show the effect of the latter on the true contact area.
This effect, undetectable for surfaces with poor spectral content, is quite strong for surfaces with rich spectra.
Numerical results are compared with analytical models and other available numerical results.
A phenomenological equation for the contact area growth is suggested with coefficients depending on the Nayak parameter. 
Using this equation, the pressure-dependent friction coefficient is deduced based on the adhesive theory of friction.
Some observations on Persson's model of rough contact, whose prediction does not depend on Nayak parameter, are reported.
Overall, the paper provides a unifying picture of rough elastic contact and clarifies discrepancies between preceding results.

\begin{flushleft}
\end{flushleft}


\section{Introduction}

Many engineering systems include components with contacts: rolling bearings, tire/road and wheel/rail, pieces assembled by bolts and rivets, gears, electric switchers, vehicle and aircraft brakes, NEMS and MEMS, etc. 
Macroscopic behavior of these components are often determined by complex electro-thermo-chemico-mechanical interactions at contact interfaces at small scales.
For many materials and structures the mechanical behavior at such scales is microstructure-dependent and near the surface can differ significantly from in-bulk behavior due to surface tension, coatings/tribo-films and oxide layers. Moreover, in most metallic devices, near-surface layers are cold hardened and recrystallized giving considerably different plastic properties. In addition, in systems with high interface stresses and/or high temperatures due to friction or fracture dissipation, near-surface microstructure may change in operation. 
An interplay between complex thermo-mechanical behavior at small scale and surface roughness coupled to other relevant physics determine the macroscopic response of the system, its life cycle and failure modes. Thus, understanding mechanical behavior of rough surfaces in mechanical contact is a key point in understanding numerous tribological systems: macroscopic friction and wear laws but also mass and heat transport along and across contacts.

Regardless the fact that numerous analytical, semi-analytical and numerical models are available nowadays, a systematic and accurate analysis of the effect of roughness parameters on contact properties is still missing. In analytical and semi-analytical models strong assumptions, introduced to make models traceable, do not allow to make quantitative predictions beyond small intervals of model validity. Multiple models relying on roughness representation by a set of individual non-interacting asperities (or with first-order interaction included in model extensions) are available and termed here asperity-based models, pioneered by works of Archard~\cite{archard1953jap} and Greenwood \& Williamson~\cite{greenwood1966prcl}, more recent models~\cite{bush1975w,thomas1999b,mccool1986w,greenwood2006w,carbone2009jmps,afferrante2012w} are based on random process theory of rough surfaces established in~\cite{longuethiggins1957rsla,nayak1971tasme}. The main drawback of these models is that elastic interaction (long-range) between asperities is missing. In reality, even if a truly rough surface could be represented by a set of asperities~\cite{greenwood2001m}, every asperity coming in contact affects vertical positions of all the surface points as $~F/r$, where $F$ is the contact reaction force and $r$ is the distance from the center of asperity. Such an interaction is not trivial to handle in analytic models, yet some models including zero-order interaction were developed~\cite{ciavarella2008w,paggi2010w}. Another drawback of asperity models is that they do not include possibility of merging contact area spots associated with different asperities, which becomes a bottle-neck limitation for surfaces with high asperity densities~\cite{greenwood2007w}. 
This drawback was also partly attenuated by including the possibility of merging and creating new macroscopic asperities~\cite{afferrante2012w}. However, to include this effect, the model should become deterministic loosing its stochastic nature, and thus it has to be integrated numerically, which requires to keep track of every individual asperity at given location and its interactions with all other asperities. 
Nevertheless, this model being sufficiently accurate is much faster than full numerical models based on boundary or finite element methods.
Finally, the curvature of each individual asperity is assumed to be constant (parabolic surface), which cannot be ensured in smooth surfaces as the sign of curvature has to change on a path connecting neighboring asperity tips, like for example in a wavy surface.

Another class of models of rough contact termed as Persson's model~\cite{persson2001jcp} is based on the evolution of contact pressure probability density (PDF) with increasing ``magnification'', the latter controls the root mean squared surface gradient. This model, being exact at full contact (which would require infinite pressure for surfaces with continuous spectra, and thus, for the height distribution defined on an infinite support) was extended to partial contacts by a boundary condition for zero-pressure. However, the diffusion equation obtained by the author for the full contact and linking the PDF of pressure with the magnification, was not rigorously proved to hold at partial contacts~\cite{manners2006w,dapp2014jpcm}. 
In addition, the author introduced a corrective function for the contact area~\cite{persson2006contact}, which should be a pressure or contact-area dependent function, to take into account differences in elastic energy for the case of partial contact compared to full contact.
It is worth also mentioning fractal models of contact~\cite{majumdar1991fractal,majumdar1990w} and discretely-continuous multi-level models with embedded long range elastic interaction between asperities~\cite{goryacheva1991ti,goryacheva2013contact,goryacheva2006ti}, which both also rely on several approximations and strong simplifications of realistic roughness.

In this light,  direct numerical simulations of rough contacts present a promising approach for solving non-linear contact problems between rough surfaces~\cite{hyun2004pre,pei2005jmps,yang2006epje,hyun2007ti,campana2008pre,paggi2010w,yastrebov2011cras,akarapu2011prl,yastrebov2012pre,putignano2012jmps,pastewka2013pre,putignano2013ti,prodanov2014tl,anciaux2009ijnme,yastrebov2015ijss}. Different methods based on boundary or finite element methods are available. On the one hand they are free of the aforementioned assumptions and limitations, but on the other hand they are subject to numerical errors due to 1) possible inaccurate treatment of mechanical behavior or contact (a prominent example would be the penalty or barrier method for constraint satisfaction); 2) discretization error and mesh convergence rate. Even though the importance of accurate discretization in contact interface was identified some time ago~\cite{yastrebov2011cras,putignano2012ijss,yastrebov2012pre,pastewka2013pre}, a true mesh convergence study was missing with notable exceptions~\cite{campana2007epl,yastrebov2011cras,putignano2013multiscale,prodanov2014tl}. In our previous work~\cite{yastrebov2015ijss} we made an effort to estimate the discretization-induced numerical error in contact area computation: it was shown that it depends not only on the ratio $\Delta x/\lambda_s$, where $\Delta x$ is the distance between grid or mesh points and $\lambda_s$ is the shortest wavelength in the surface spectrum, but that the error also depends on the ratio $\lambda_l/L$, where $L$ is the length of the simulation box and $\lambda_l$ is the longest wavelength in the surface spectrum for spectra without plateau. The first ratio $\Delta x/\lambda_s$ should tend to zero to capture accurately the mechanical behavior of each asperity, thus at grid size the surface should remain smooth contrary to what was done in early numerical studies~\cite{hyun2004pre,pei2005jmps,paggi2010w}. The second ratio $\lambda_l/L$ determines representativity of the surface~\cite{yastrebov2012pre,yastrebov2015ijss}, i.e. the proximity of the surface height distribution to a Gaussian one, so the representative surface limit can be reached for $\lambda_l/L \to 0$, which is also a numerically unreachable limit. High values of this ratio (of the order of unity) thus require an extensive statistical study and averaging over many roughness realizations to extract accurate mean values, similar to what is done for homogenization of mechanical properties~\cite{kanit2003determination}.
However, in many studies $\lambda_l/L = 1$, i.e. the longest wavelength in the surface spectrum is equivalent to the simulation box (which is often assumed to be periodic), which implies a considerable dispersion in results and requires many simulations to obtain a statistically meaningful mean behavior.
In~\cite{yastrebov2015ijss} it was also shown that when the ratio $\lambda_l/L$ decreases the true contact area increases: this scenario appeared reasonable to the authors, as it was argued~\cite{yastrebov2012pre} that the periodicity coupled to long range elastic interactions and non-representativity of the system may affect significantly the results.
What was inconsistent in~\cite{yastrebov2015ijss} is that no convergence was observed even for quite representative surfaces with $\lambda_l/L = 1/16$. In other words, the surface becomes more and more representative and isotropic but the true contact area continues to increase when $\lambda_l/L$ decreases further. This spurious behavior was accounted for discretization errors, and thus all numerical results contaminated by this error could not be considered ultimately correct.

In our recent study~\cite{yastrebov2016w} we performed an accurately designed mesh convergence test and suggested a specific technique to correct the measure of the contact area. The technique is based on the error estimation suggested in~\cite{yastrebov2015ijss} and some geometric arguments. The resulting equation for the true contact area includes the measured contact area in simulations $A_{\mbox{\tiny sim}}$ and the contact perimeter $S$, i.e. the number of switches between contact and non-contact points along vertical and horizontal lines of the grid, scaled with the grid size $\Delta x$. Then the true contact area can be estimated as
\be
  A_* = A_{\mbox{\tiny sim}} - \frac{\pi - 1 + \ln2}{24} S\Delta x.
  \label{eq:cor_area}
\ee
This form provides a very accurate estimation of the contact area: even for poorly discretized surfaces $\Delta x/\lambda_s = 1/2$ this equation provides as accurate estimation as 
a simulation carried out on a much finer mesh $\Delta x/\lambda_s = 1/32$ for unchanged spectral content of the surface.

With Eq.~\eqref{eq:cor_area} in hand, in this paper we carry out numerous numerical simulation and analyze the results to obtain contact area evolution under increasing pressure with unprecedented accuracy.
Based on these results, several important conclusions are made concerning the role of the two roughness parameters: first, the spectral breadth (so-called Nayak parameter) and, second, the fractal dimension or equivalently the Hurst exponent. Both parameters are strongly linked, but surprisingly the role of the former one has not yet been thoroughly studied in the framework of full numerical models.

The paper is organized as follows: in Section~\ref{sec:method} the numerical methods and roughness models are briefly outlined, in Section~\ref{sec:results} numerical results are presented, namely the contact perimeter, contact area, inverse mean pressure and derivative of the contact area with respect to the nominal pressure. 
A simple phenomenological equation is suggested for the contact area evolution with universal coefficients depending on the Nayak parameter.
Finally, the role of Nayak parameter in contact between rough surfaces is put in evidence.
Implications of our results for Persson's model of contact, which does not depend on Nayak parameter, is also discussed.
The results are summarized and interpreted in Section~\ref{sec:discusion}.
In Appendix some additional plots and tables are presented. All numerical data can be found in Supplemental material~\cite{supplemental}.

\begin{figure}[htb!]
  \includegraphics[width=1\textwidth]{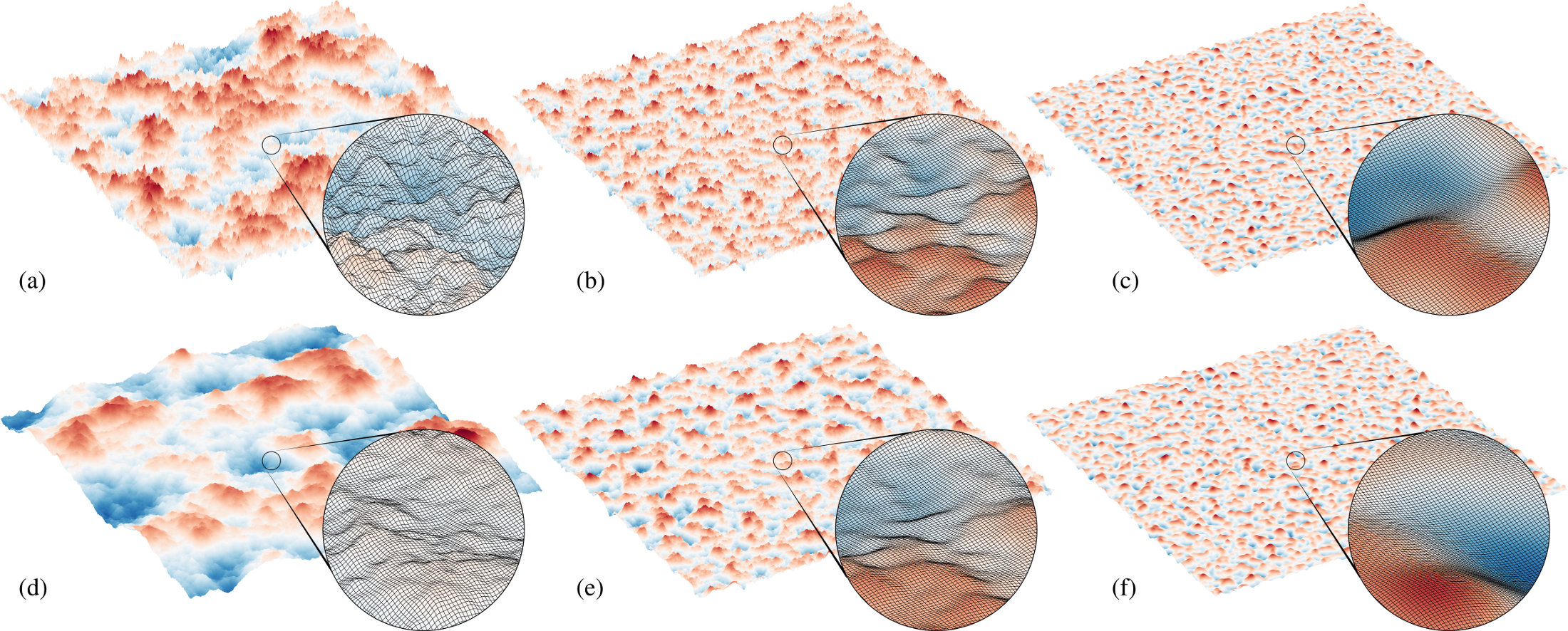}
  \caption{\label{fig:rough}Synthetic rough surfaces with $2048\times2048$ points and zooms on small portions showing $\approx50\times50$ grids:  
  (a) $L/\lambda_l=2$, $L/\lambda_s=512$, $H=0.4$, $\alpha \approx 67.8$; 
  (b) $L/\lambda_l=8$, $L/\lambda_s=256$, $H=0.4$, $\alpha \approx 12.9$; 
  (c) $L/\lambda_l=16$, $L/\lambda_s=64$, $H=0.4$, $\alpha \approx 2.6$; 
  (d) $L/\lambda_l=2$, $L/\lambda_s=512$, $H=0.8$, $\alpha \approx 814.3$; 
  (e) $L/\lambda_l=8$, $L/\lambda_s=256$, $H=0.8$, $\alpha \approx 27.7$; 
  (f) $L/\lambda_l=16$, $L/\lambda_s=64$, $H=0.8$, $\alpha \approx 2.7$. 
  }
\end{figure}

\section{Methods \label{sec:method}}

\subsection{Synthetic rough surfaces}

To generate self-affine rough surfaces, we use a filtering-in-Fourier-space technique~\cite{hu1992ijmtm}. 
Four parameters and a seed for a random number generator fully determine the roughness: $k_l = 2\pi/\lambda_l$ and $k_s = 2\pi/\lambda_s$ are wavenumbers determining the lower and higher cutoffs in the discrete spectrum. We deal with a periodic square surface of side $L$, thus the shortest and the longest cutoff wavelengths should be defined as $\lambda_s = L/n_s$,$\lambda_l = L/n_l$, where $n_s,n_l \in \mathbb N$. The third parameter $\Phi_0$ is the amplitude scaling parameter and $H$ is the Hurst exponent.
The method consists in the following. We generate a white noise $w_{ij} = w(x_i,y_j)$ on a grid with $N\times N$ points, such that $\langle w \rangle = 0$ and $\langle w^2\rangle = \Phi_0$, where $\langle \bullet \rangle$ denotes the mean value over the domain:
$$
\langle x \rangle = \frac{1}{N^2}\sum\limits_{i=0}^{N-1}\sum\limits_{j=0}^{N-1} x_{ij}.
$$
It is transformed in Fourier space (Discrete Fourier Transform):  
$$
\hat w_{mn} = \sum\limits_{i=0}^{N-1}\sum\limits_{j=0}^{N-1} w(x_i,y_i) \exp\left[- 2\pi\mathrm{i}(m x_i + n y_j)/L\right],
$$
so that $\langle \hat w \hat w^* \rangle = \langle w^2\rangle = \Phi_0$, where $\hat w^*$ denotes complex conjugate of $\hat w$.
In Fourier space we create a power-law decaying filter which retains wavenumbers only in the interval $k_l\le k \le k_s$ and scales their amplitude according to a decaying power-law:
\bes
\hat f_{ij} = \hat f(K_i,K_j) = \begin{cases}
			      \left[\frac{K_i^2+K_j^2}{k_l^2}\right]^{-(1+H)/2},\mbox{ for }1 \le \frac{\sqrt{K_i^2 + K_j^2}}{k_l} \le \zeta\\
                                 0, \mbox{ elsewhere,}
                                \end{cases}
\ees
where $K_i = (s+1)\pi/L-2\pi s i/L$, $K_j = (t+1)\pi/L-2\pi t j/L$, for $s,t\in\{-1,1\}$, $\zeta = k_s/k_l$ is referred as magnification in Persson's model~\cite{persson2001jcp}, and the parameter $H$ is the Hurst exponent determining the rate of decay, which in fractal description of 2D surfaces is linked to the fractal dimension $D_f$ as $H=3-D_f$.
Note that imaginary part of the filter is everywhere zero $\Im{f} = 0$.
Next, we take a product of the white noise with the filter, which in Fourier space results in
$$
  \hat z_{ij} = \hat z(k_x,k_y) = \Re(\hat f_{ij})\left[\Re(\hat w_{ij}) + \mathrm{i}\, \Im(\hat w_{ij})\right].
$$
The field is transformed back to the real space to provide the targeted rough surface
$$
z(x_i,y_j) = \sum\limits_{n=0}^{N-1}\sum\limits_{m=0}^{N-1} \hat z_{nm} \exp[\mathrm{i}2\pi(n x_i + m y_j)/L].
$$ 
Note that the imaginary part of $z$ is zero.
The discrete auto-correlation function is defined for the periodic $L\times L$ surface as
$$
R(\Delta x,\Delta y) = \frac{1}{N^2} \sum\limits_{i=0}^{N-1} \sum\limits_{j=0}^{N-1} z(x_i+\Delta x,y_j+\Delta y) z(x_i,y_j),
$$
and finally the power spectral density (PSD) of the roughness defined as the discrete Fourier transform (DFT) of the auto-correlation function, is given by:
$$
  \Phi(k_x,k_y) = \hat R(k_x,k_y) = \mathrm{DFT}\left[z(x+\Delta x,y+\Delta y) * z(x,y)\right], 
$$
which according to the convolution theorem can be found by point-wise product of the Fourier transform of $z$ as
$$
\Phi(k_x,k_y) = \hat z(k_x,k_y) \hat z^*(k_x,k_y) = \hat f^2(k_x,k_y) \hat w(k_x,k_y)\hat w^*(k_x,k_y) = \hat f^2(k_x,k_y) \hat w^2(k_x,k_y).
$$
Thus the PSD in average decays proportionally to $\hat f^2(k_x,k_y)$, because the random value $\hat w^2(k_x,k_y)$ in average does not depend on $k_x,k_y$.
It is also clear that for random realizations of the white noise the following statement is ensured:
$$
 \langle\Phi(k_x,k_y)\rangle = \langle \hat w^2(k_x,k_y) \rangle \hat f^2(k_x,k_y)  = 
			      \begin{cases}
			      \Phi_0\left[\frac{\sqrt{K_x^2+K_y^2}}{k_l}\right]^{-2(1+H)},\mbox{ for }1 \le \frac{\sqrt{K_x^2 + K_y^2}}{k_l} \le \zeta\\
                                 0, \mbox{ elsewhere,}
                                \end{cases} 
$$
here $\langle \bullet \rangle$ denotes the average over multiple realizations of the white noise.
It is the targeted form of the PSD, which follows the power-law-decay with the wavenumber with the exponent $-2(1+H)$.
If the resulting surface is isotropic (spectral content is independent in statistical sense of the profile orientation on the surface), and it is supposed to be so if the filter is axi-symmetric and the $k_l$ parameter is big enough, then the PSD can be represented as a function of a single wavenumber $K = \sqrt{K_x^2+K_y^2}$: 
\bes
\langle \Phi(K) \rangle = \begin{cases}
	      \Phi_0 (K/k_l)^{-2(1+H)},&\mbox{ if } 1 \le K/k_l \le \zeta\\
	      0,&\mbox{ otherwise}.
            \end{cases}
\ees
Spectral moments $m_{pq}$ for a surface with a discrete spectrum are computed as follows:
\be
  m_{pq} = \left[\frac{2\pi}{L}\right]^{p+q}\sum\limits_{i = 0}^{N-1}\sum\limits_{j=0}^{N-1}  i^p j^q\; \Phi(2\pi i/L,2\pi j/L),
  \label{eq:spectral_moments}
\ee
under the assumption of isotropic surface and for a broad spectrum, spectral moments can be approximated by an integral~\cite{yastrebov2015ijss}:
\be
 m_{0p} \approx m_{p0} \approx \Phi_0\int\limits_{k_l}^{k_s}\int\limits_{0}^{2\pi} \left[k\cos(\varphi)\right]^{p} (k/k_l)^{-2(1+H)}\,kdkd\varphi 
 = \Phi_0 k_l^{p+2} \frac{\zeta^{p-2H}-1}{p-2H} T(p),
\label{eq:a:mq0}
\ee
where $T(p)$
$$
T(p) = \int\limits_{0}^{2\pi} \cos^p(\varphi)d\varphi
= \begin{cases} 
  2\pi,&\mbox{ if }p=0;\\ 
  \pi,&\mbox{ if }p=2;\\ 
  3\pi/4,&\mbox{ if }p=4.\\
 \end{cases}
$$
Isotropy of the surface, in the first approximation, can be verified by the following relationships between moments:
$m_2 = m_{20} = m_{02}$, $m_4 = 3m_{22} = m_{40} = m_{04}$. For a surface with a discrete spectrum, these equalities are satisfied approximately, thus
to even out errors one can rather use the following expressions for the spectral moments:
\be
  m_2 = \frac{m_{20} + m_{02}}{2},\quad m_4 = \frac{m_{40} + 3m_{22} + m_{04}}{3}.
  \label{eq:av_moments}
\ee
The Nayak parameter $\alpha\in[1.5,\infty]$ introduced in~\cite{nayak1971tasme} determines the breadth of the surface spectrum and is given by the following expression:
$$
  \alpha = m_0 m_4/m_2^2
$$
Below, we recall some basic definitions and relationships between spectral moments and the variances of surface heights, of surface gradient and of surface curvature:
$$
  \left\langle (z-\langle z \rangle)^2\right\rangle = m_0,\quad \left\langle (\nabla z-\langle \nabla z \rangle)^2\right\rangle = 2m_2,\qquad \left\langle (\nabla\cdot\nabla z-\langle \nabla\cdot\nabla z \rangle)^2\right\rangle = m_4,
$$
where mean values are computed over a surface period $L\times L$, i.e. $\langle\bullet\rangle = \frac{1}{L^2}\int\limits_0^L\int\limits_0^L \bullet \,dxdy$.
It is important to note that in the case of a discrete spectrum, these relationships are not exact and are strongly dependent on the surface discretization and on the way how gradients and Laplacians are computed~\cite{greenwood1984unified,paggi2010w}. Thus to obtain a discretization-independent measurement of Nayak parameter and of the root mean squared (rms) gradient, which is abbreviated hereinafter as $\sqrt{\langle|\nabla z|^2\rangle}$, assuming that $\langle\nabla z\rangle = \boldsymbol 0$, we evaluate them through spectral moments~\eqref{eq:spectral_moments} and use average value~\eqref{eq:av_moments}: $\sqrt{\langle|\nabla z|^2\rangle} = \sqrt{m_{02}+m_{20}}$.

The following combinations of parameters were considered in this study:
$L/\lambda_l = Lk_l/2\pi = \{2, 4, 8, 16\}$, $L/\lambda_s = Lk_s/2\pi = \{16, 32, 64, 128, 256, 512\}$, $H=\{0.4,0.8\}$. For every combination of parameters (46 combinations excluding cases $k_l = k_s$) ten different roughness were generated. For all surfaces the rms surface gradient is kept constant $\sqrt{\langle|\nabla z|^2\rangle}=1$. All generated rough surfaces are periodic and are constructed on the grid $N\times N = 2048\times 2048$.
Some examples of generated rough surfaces, which are used in numerical simulations, are depicted in Fig.~\ref{fig:rough}. 

\begin{figure}[htb!]
 \includegraphics[width=1\textwidth]{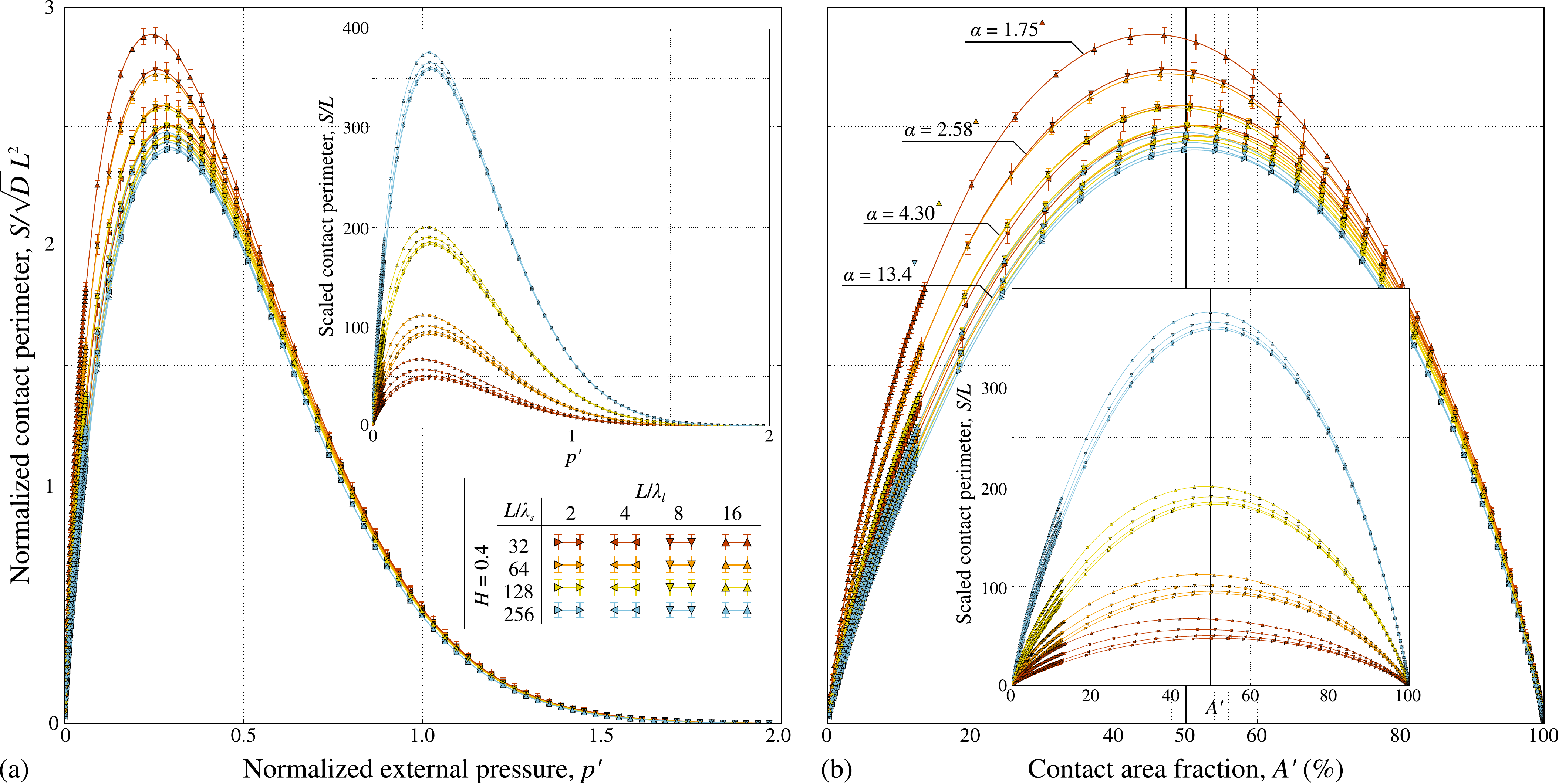}
 \caption{\label{fig:perimeter04}Evolution of the normalized contact perimeter $S/\sqrt{D}L^2$ for $H=0.4$ and for different $L/\lambda_l$ and $L/\lambda_s$ with respect to (a) normalized nominal pressure $p'$, Eq.~\eqref{eq:norm_perim}, (b) real contact area fraction (corrected contact area~\eqref{eq:cor_area} is used). Every point represents the mean value computed over simulations carried out for 10 different realizations of roughness. Hereinafter the color corresponds to $k_s$ cutoff and the shape of the insertion corresponds to $k_l$ cutoff.
 Non-normalized perimeter evolution $S/L$ is shown in the insets.
 }
\end{figure}

\begin{figure}[htb!]
 \includegraphics[width=1\textwidth]{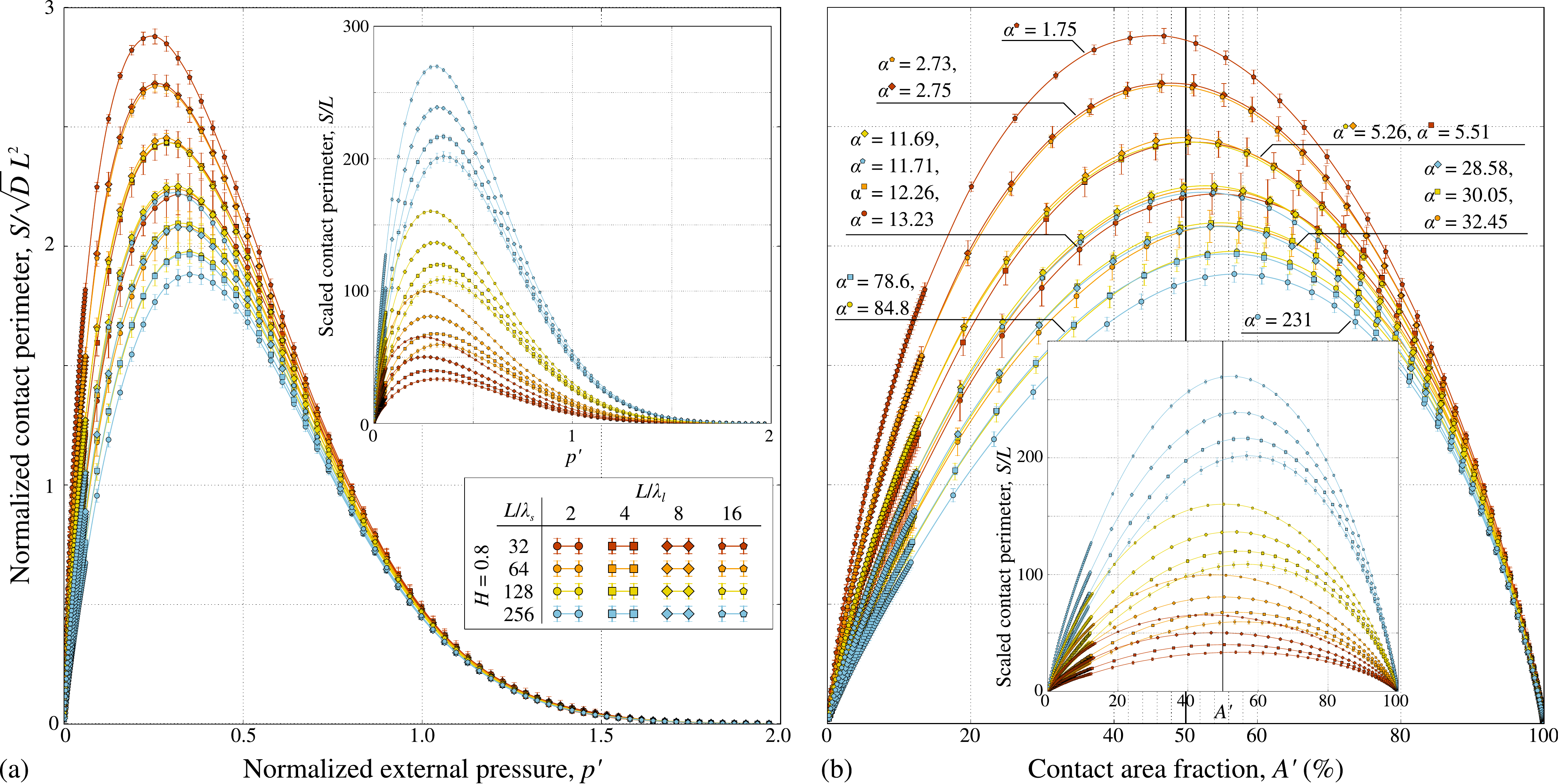}
 \caption{\label{fig:perimeter08}Evolution of the normalized contact perimeter $S/\sqrt{D}L^2$ for $H=0.8$ and for different $L/\lambda_l$ and $L/\lambda_s$ with respect to (a) normalized nominal pressure $p'$, Eq.~\eqref{eq:norm_perim}, (b) real contact area fraction (corrected contact area~\eqref{eq:cor_area} is used). Every point represents the mean value computed over simulations carried out for 10 different realizations of roughness. Non-normalized perimeter evolution $S/L$ is shown in the insets.}
\end{figure}

\subsection{Comments on the roughness and contact models' validity}

Note that contrary to the zeroth moment (or the variance of height), spectral moments $m_2$ and $m_4$ diverge when $Lk_s \to \infty$. 
Thus their determination from experimental roughness measurements depend on the magnification and measurement sampling, i.e. on the grid step.
Roughly speaking, the value of $m_2$ is measurement-dependent, and since $m_2$ is the main scaling parameter in rough contact models of elastic contact, the output of these models is also measurement-dependent. 
Since many surfaces demonstrate self-affine scaling and power-law decrease of the power spectral density down to interatomic distances~\cite{krim1995ijmpb}, where continuum mechanics and especially contact interaction~\cite{bhushan1995n,robbins2005n,bhushan2008b} are rigorously not applicable, then the study of elastic contact becomes questionable. 
Moreover, at atomic scale the definition of contact becomes ambiguous as solid-solid interaction is determined by interactions with infinite support: short-range interaction in absence of Coulomb electrostatic forces, and long-range interaction in their presence.
Therefore to use all existing models of elastic rough contact one needs to assume that the surface is smooth below a certain wavelength $\lambda_s$, which certainly is quite a rough approximation. A different approach would be to consider that elastic interaction does not longer holds at shorter wavelengths and other scale-related interface features are dominant: atomic or capillary adhesion, surface energy, etc.
Moreover, the cutoff problem is easy to address if elasto-plastic material behavior is considered for which too sharp asperities coming in contact are readily flattened out in fully plastic regime, for which the contact pressure saturates at material hardness~\cite{bowden2001b}. However, not all materials exhibit similar-to-metals plastic behavior (e.g. pressure dependent plasticity in rocks or amorphous solids), moreover at smaller scales hardness becomes size-dependent~\cite{nix1998jmps,swadener2002jmps}, which might bias the conclusions of models based on macroscopic hardness.
In summary, for the time being the question of physically motivated high-frequency cutoff remains open and will not be addressed in this study.

\subsection{Mechanical contact between rough surfaces}

We use a spectral boundary element method proposed in~\cite{stanley1997}, which is rather similar to~\cite{johnson1985ijms}, the method is based on the Kalker's formulation of the minimization problem under constraints~\cite{kalker1972jem}. We solve the normal contact problem between a synthetic rough surface, assumed to be rigid and a linearly elastic half-space. The roughness of the rigid surface should be considered as an effective roughness of two contacting solids $z(x,y) = z_1(x,y) - z_2(x,y)$. The slopes are assumed to be very small: $|\nabla z| \ll 1$ so that tangential displacements are negligible and the classical approach~\cite{johnson1987b}, where the loads are assumed to be applied on a flat surface, is ensured.
Only the normal, frictionless and non-adhesive contact is considered. The only material parameter used, is the effective elastic modulus given by:
$$
  E^* = \frac{E_1E_2}{(1-\nu_1^2)E_2+(1-\nu_2^2)E_1},
$$
where $E_i,\nu_i$ are Young's modulus and Poisson's ratio of material $i=1,2$, respectively.

The normalized nominal pressure applied at infinity is computed as
$$
p' = \frac{p_0}{E^*\sqrt{\langle|\nabla z^2|\rangle}}
$$
and is increased from zero up to $p' = 2$ in 120 load steps; to ensure accurate results at small contacts, 60 load steps are used to reach $p' = 0.06$.
We store spectral moments of the surface and for every load step we store the perimeter of contact clusters $S$, and the contact area $A_{\mbox{\tiny sim}}$. The perimeter is computed as a number of switches from contact to non-contact and vice versa along all vertical and horizontal lines on the grid multiplied by the discretization step $\Delta x$, the contact area is computed as the number of grid points in contact multiplied by the area corresponding to every point $\Delta x^2$. In the paper we present the true contact area fraction given by the ratio $A' = A/A_0$, where $A_0 = L^2 = N^2\Delta x^2$. 

\section{Results and Discussions~\label{sec:results}}


\subsection{Contact perimeter}

Since the contact perimeter plays an important role in the accurate estimation of the true contact area~\cite{yastrebov2016w}, it is of interest to know how it changes under increasing load.
Evolution of the contact perimeter (normalized and non-normalized data) with respect to the true contact area and the nominal pressure is presented in Figs.~\ref{fig:perimeter04},\ref{fig:perimeter08} for $H=0.4,0.8$, respectively.
The normalization $S' = S/(\sqrt{D}L^2)$ is found from the following consideration: if the contact is formed by identical circles of radius $r$, then the total area is given by $A \sim \pi D L^2 r^2 $, where $L$ is the side length of the considered region and $D$ is the total asperity density given in Nayak's paper~\cite{nayak1971tasme}: which is given by
$$
  D = \frac{\sqrt3}{18\pi}\frac{m_4}{m_2},
$$
which should be proportional to the density of contacting asperities $D_c$.
At the same time the total perimeter of contact spots is given by
$S \sim 2\pi D L^2 r$. Expressing $r$ through $A$ and $D$, we obtain that 
\be
S \sim L\sqrt{AD}.
\label{eq:perim_scaling}
\ee
It means that for equivalent contact area $A$, which depends rather weakly on parameters $k_l$ and $k_s$, the perimeter scales as $S\sim L^2\sqrt{D}$.
Then expressing asperity density through $k_l,k_s$ and $H$ using Eq.~\eqref{eq:a:mq0}, we finally obtain the following normalization:
\be
  S' = S/(\sqrt{D}L^2) = \frac{S}{k_l L^2}\sqrt{\frac{24\pi}{\sqrt3}\cdot\frac{2-H}{1-H}\cdot\frac{\zeta^{2-2H}-1}{\zeta^{4-2H}-1}}.
  \label{eq:norm_perim}  
\ee

As seen in Figs.~\ref{fig:perimeter04},\ref{fig:perimeter08}, the perimeter is not symmetric with respect to $A' = A/A_0 = 50$ \%, demonstrating the asymmetry in evolution of contact and non-contact clusters (see also~\cite{yastrebov2014tl}). 
Interestingly, for the chosen normalization, the slope of the perimeter approaching full contact is almost the same for all roughnesses [Figs.~\ref{fig:perimeter04}(b),\ref{fig:perimeter08}(b)].
On the contrary, for all cutoffs the normalized perimeter evolves differently near zero contact. 
The maximum of the perimeter, which corresponds to the maximal \emph{absolute} error\footnote{Note that the maximal \emph{relative} error is always reached at minimal computed contact area, see Fig.~\cite[Fig. 12]{yastrebov2015ijss}} in contact area computation, can be easily identified in Figs.~\ref{fig:perimeter04},\ref{fig:perimeter08}. 
Note that the normalized perimeter decreases with Nayak parameter $\alpha$, whereas the area corresponding to the peak perimeter increases with increasing $\alpha$. 
For surfaces with high enough $\alpha$ the contact area corresponding to the maximum of the perimeter seems to converge toward $A' = A/A_0 \approx 57$ \% [see Fig.~\ref{fig:perimeter08}(b)]. 
It could seem to be meaningful to link this contact area with the percolation area of contact clusters, which corresponds to the appearance of an infinite contact cluster.
However, this value appears to be considerably higher than the percolation limits found i) for a bi-wavy surface $A'_p \approx 40.2$ \%~\cite{yastrebov2014tl} and ii) for random rough surfaces $A'_p \approx 40$ \%~\cite{aharony2003book}, $A_p \approx 42.5$ \%~\cite{dapp2012prl},  $A'_p \approx 35.5$ \%~\cite{putignano2013multiscale}, a broader range of percolation limit was reported in~\cite{sahlin2008lubrication}.

\begin{figure}[htb!]
 \centering\includegraphics[width=1\textwidth]{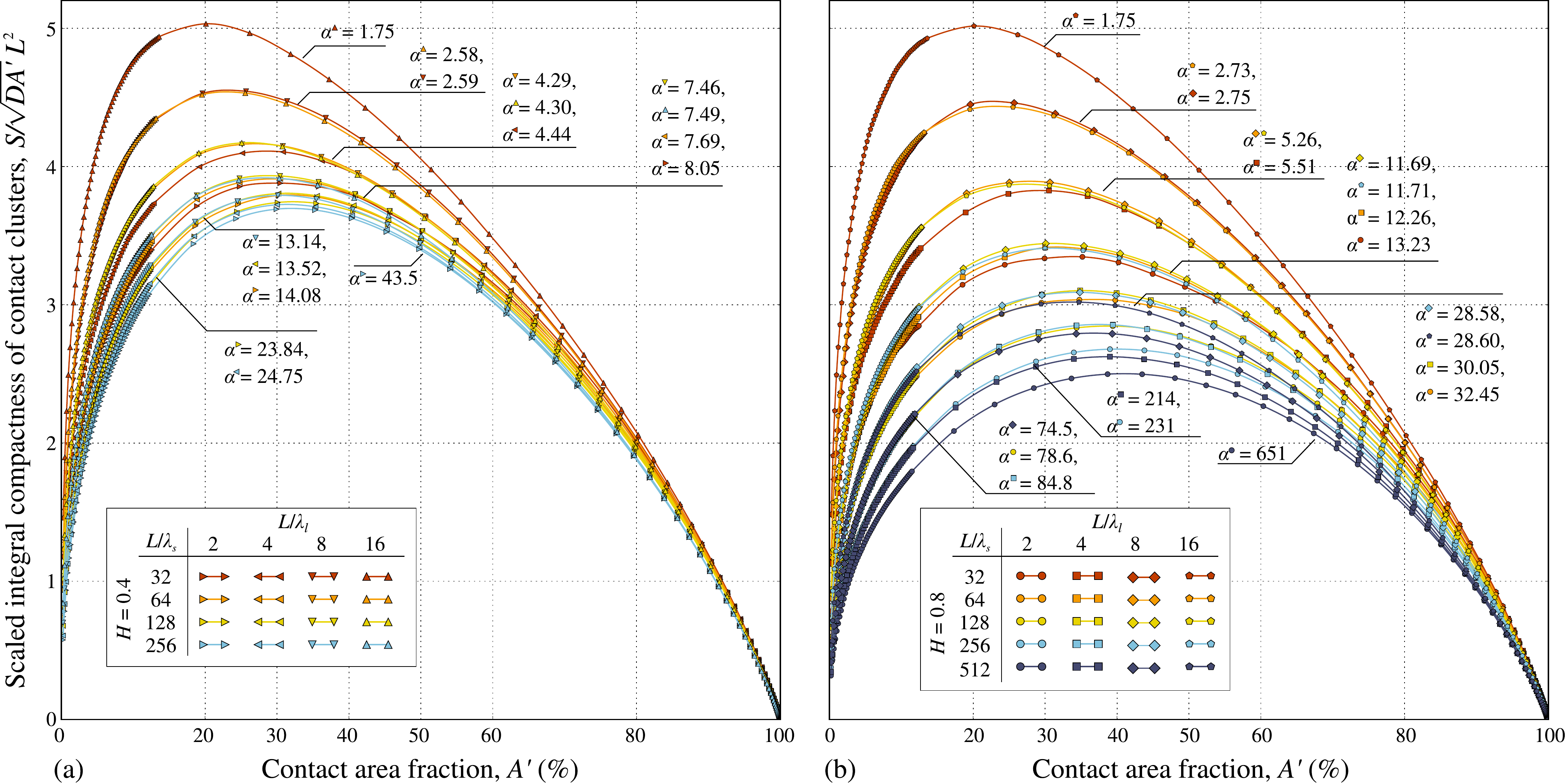}
 \caption{\label{fig:compactness}Evolution of the normalized contact perimeter divided by square root of the contact area (integral compactness, see~\cite{yastrebov2014tl}) $S/\sqrt{DA'}L^2$ with respect to the real contact area (corrected contact area is used) computed for (a) $H=0.4$, (b) $H=0.8$ and for different $L/\lambda_l$ and $L/\lambda_s$.}
\end{figure}

Another remark on the perimeter deals with a dimensionless characteristic of the shape, compactness~\cite{yastrebov2014tl}, which can be defined as a ratio of the perimeter to square root of the contact area: $C=S/\sqrt{A}$. In the simplest case of identical circular contact spots, following~\eqref{eq:perim_scaling} the compactness is a constant proportional to the square root of asperity density $S/\sqrt{A} \sim L\sqrt{D}$, however, in a realistic rough contact it is not the case.
Evolution of the compactness for different roughnesses is shown in Fig.~\ref{fig:compactness} for both $H=0.4,0.8$. 
Note that if normalized perimeter is used to determine the compactness, as it is done in Fig.~\ref{fig:compactness},
$C' = S/\sqrt{DA'}L^2$, then the normalization $DA'$ can be interpreted as the density of contacting asperities, whereas $D$ is the total density of asperities.
Interestingly, the area value corresponding to the maximum of the compactness also increases with the Nayak parameter $\alpha$ and saturates at $A/A_0 \approx 42$ \%, which is intriguingly close to the above-mentioned percolation limits found by other authors, for instance $A_p \approx 42.5$ \% from~\cite{dapp2012prl}.
In this paper we limit ourselves to this short remark on the link found between the percolation limit and the contact area at which the maximal value of compactness is reached. 
A further study is needed to confirm this correlation and to find a rigorous explanation for it.

\subsection{Contact area}

\begin{figure}[htb!]
 \includegraphics[width=1\textwidth]{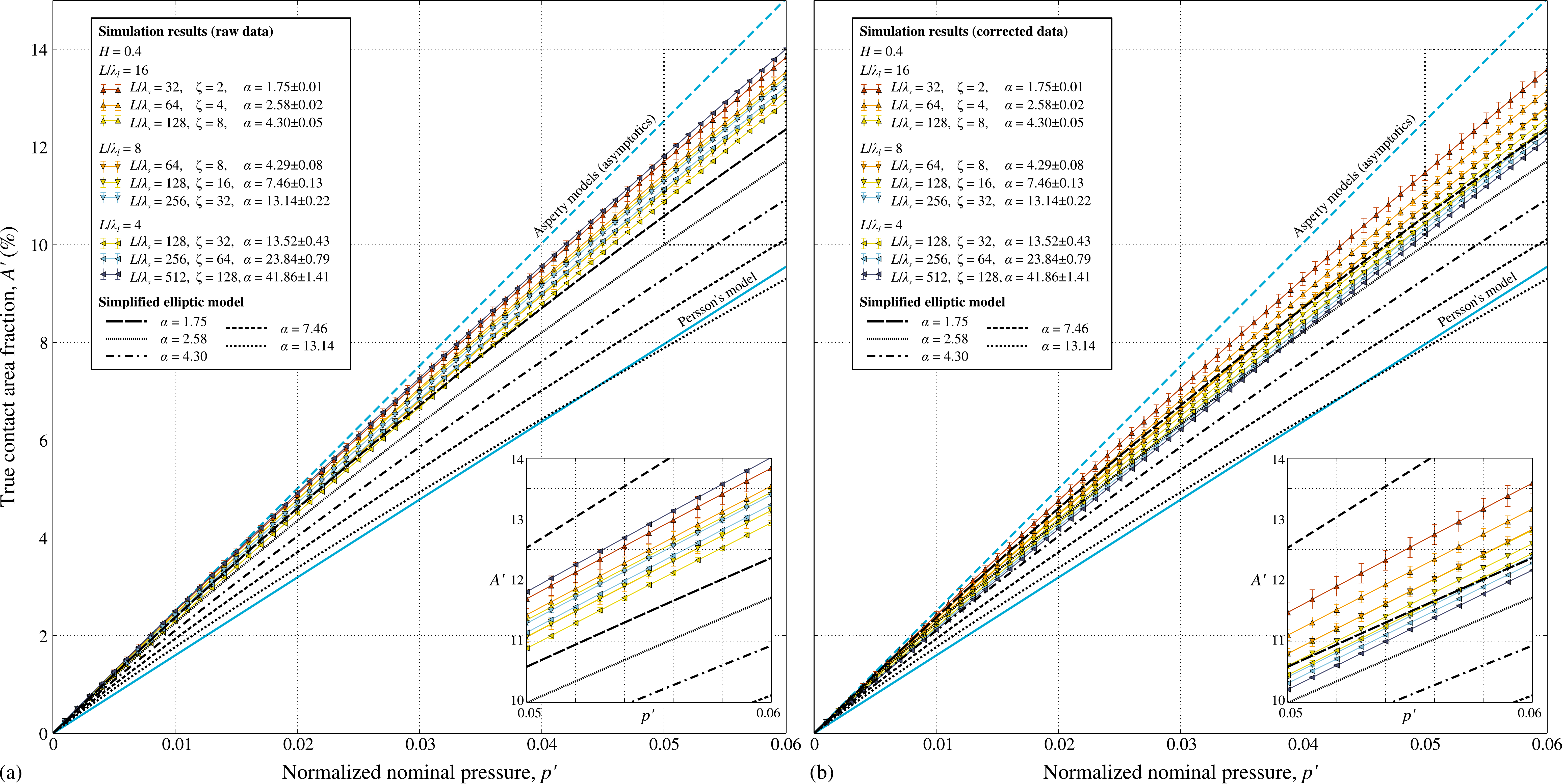}
 \caption{\label{fig:area_04}True contact area evolution with the normalized nominal pressure $A'(p')$ computed for $H=0.4$: (a) raw simulation data, (b) corrected simulation data obtained using~\eqref{eq:cor_area}.
 Simulation results (lines with points) are compared with analytic models: Persson's model (solid light line), asymptotic linear solution of asperity models~\cite{bush1975w,carbone2008jmps} (dashed light line) and the Greenwood's simplified elliptic model~\cite{greenwood2006w} integrated for $\alpha = 1.75,\;2.58,\;4.30,\;7.46,\;13.14$ (black dashed and dotted lines). A zoomed region is shown in the inset.
 }
\end{figure}

\begin{figure}[htb!]
 \includegraphics[width=1\textwidth]{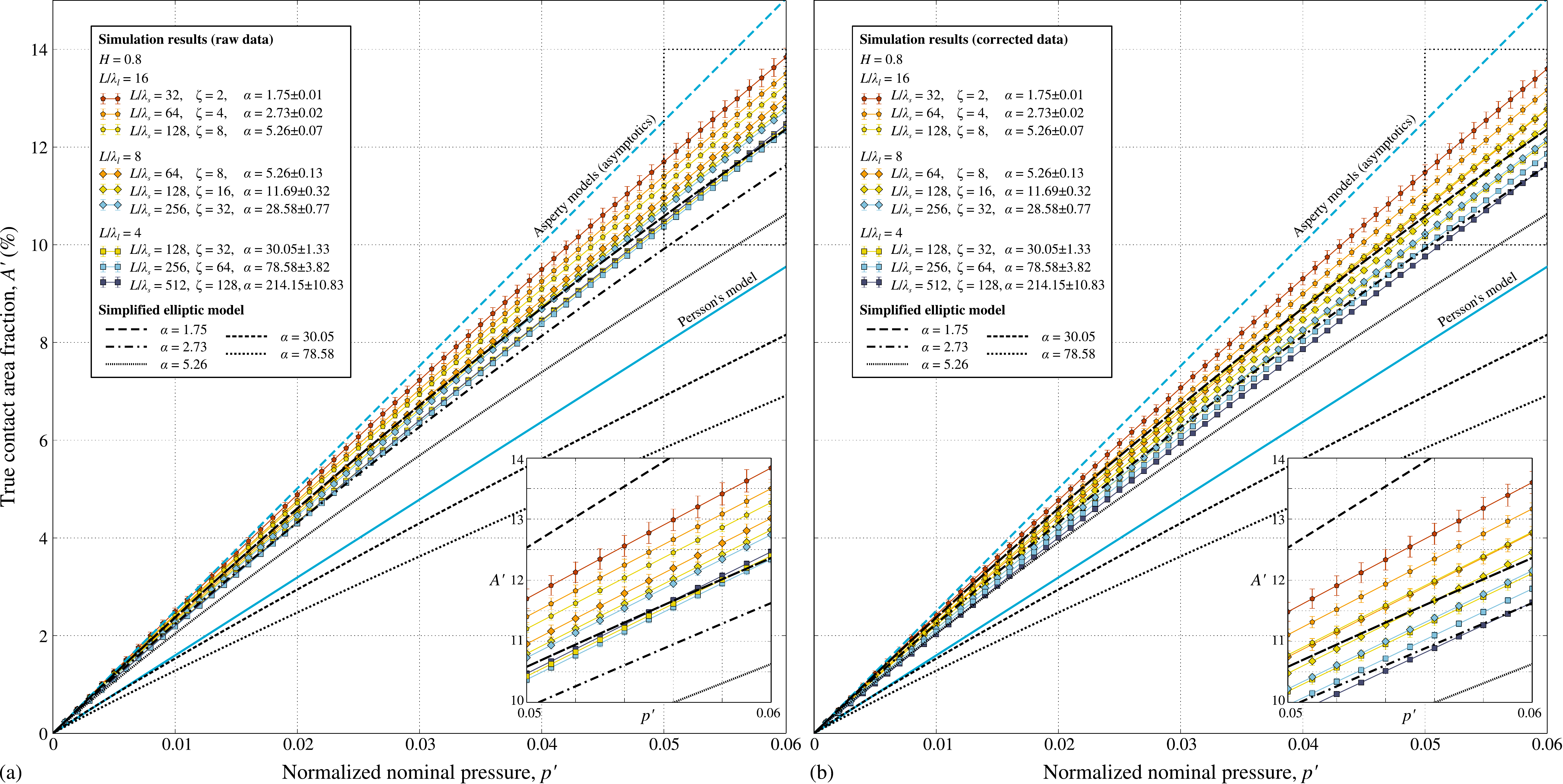}
 \caption{\label{fig:area_08}True contact area evolution with the normalized nominal pressure $A'(p')$ computed for $H=0.8$: (a) raw simulation data, (b) corrected simulation data obtained using~\eqref{eq:cor_area}.
 Simulation results (lines with points) are compared with analytic models: Persson's model (solid light line), asymptotic linear solution of asperity models~\cite{bush1975w,carbone2008jmps} (dashed light line) and the Greenwood's simplified elliptic model~\cite{greenwood2006w} integrated for $\alpha = 1.75,\;2.73,\;5.26,\;30.05,\;78.58$ (black dashed and dotted lines). A zoomed region is shown in the inset.
 }
\end{figure}

Here, we present numerical results of the contact area evolution computed for surfaces with different spectra. 
Some numerical data are presented in Tables~\ref{tab:3},\ref{tab:4} in the Appendix, 
all data can be found in supplementary material~\cite{supplemental}, which we provide for reader's convenience.
All results were post-processed using the suggested area-correcting technique Eq.~\eqref{eq:cor_area} (see~\cite{yastrebov2016w}). In Figs.~\ref{fig:area_04},\ref{fig:area_08} the true contact area evolution with normalized nominal pressure is shown for $H=0.4,0.8$, respectively.

Looking on raw non-corrected results Figs.~\ref{fig:area_04}(a),\ref{fig:area_08}(a), it can be concluded that Nayak parameter $\alpha$ has a non-monotonous effect on the contact area, which contradicts the prediction of all asperity-based models~\cite{carbone2008jmps} that higher $\alpha$ results in a smaller contact area for the same normalized pressure $p'$.
Compare, for example, in Fig.~\ref{fig:area_04} results obtained for $\alpha=1.75$, $\alpha=13.52$ and $\alpha = 41.86$, the corresponding contact areas found at $p'\approx0.06$ first decreases and next increases with increasing $\alpha$.
It also could seem from the raw data that the lower cutoff $L/\lambda_l$ plays an important role in contact area determination.
Compare, for example, in Fig.~\ref{fig:area_08}(a) surfaces $L/\lambda_l=16,L/\lambda_s=128$ and $L/\lambda_l=8,L/\lambda_s=64$ which have the same $\alpha = 5.26$ and the same magnification $\zeta = k_s/k_l$, but different lower frequency cutoff $\lambda_l$.

After the correction (Figs.~\ref{fig:area_04}(b),\ref{fig:area_08}(b)], clearly, this spurious dependence of results on the lower cutoff in the surface spectrum $\lambda_l$ and on the Nayak parameter disappears. The former was already identified in~\cite{yastrebov2015ijss} (see also Fig.~\ref{fig:area_ijss} in Appendix for post-processed results of~\cite{yastrebov2015ijss}). In corrected results, close values of Nayak parameter result in close values of the contact area: 
compare, for example, in Fig.~\ref{fig:area_08}(a,b) $\alpha = 5.26$ for $L/\lambda_l=16,L/\lambda_s=128$ and $L/\lambda_l=8,L/\lambda_s=64$,
or $\alpha = 28.6,L/\lambda_l=8,L/\lambda_s=256$ and $\alpha=30.05,L/\lambda_l=4,L/\lambda_s=128$.
Complementary figures~\ref{fig:area04},\ref{fig:area08} demonstrating the effect of the correction can be found in Appendix.
Simulation results are also compared with analytical models: asymptotic solution of Nayak's based asperity models $A' = \sqrt{2\pi} p'$~\cite{bush1975w,mccool1986w,greenwood2006w,carbone2009jmps,thomas1999b}, original Persson's model $A' = \erf(p'\sqrt2)$~\cite{persson2001jcp}, and numerically integrated Greenwood's simplified elliptic model~\cite{greenwood2006w} for relevant values of Nayak parameter. We recall that in the later model the contact area and the nominal pressure are defined as functions of dimensionless separation between the effective rough surface and a rigid flat $\hat s = s/\!\!\sqrt{m_0}$ as follows:
\be
 A' = \frac{\sqrt\alpha}{6}\int\limits_{\hat g=0}^\infty\int\limits_{\xi = \hat s}^\infty P(\xi,\hat g)\frac{(\xi-\hat s)}{\hat g}\,d\xi \,d\hat g,
  \label{eq:intA}
\ee
\be
 p_0 = E^*\sqrt{m_2/\pi}\frac{2\alpha^{3/4}}{9\sqrt\pi}\int\limits_{\hat g=0}^\infty\int\limits_{\xi = \hat s}^\infty P(\xi,\hat g)\frac{(\xi-\hat s)^{3/2}}{\sqrt{\hat g}}\,d\xi \,d\hat g,
  \label{eq:intF}
\ee
where $\hat g = g/\!\!\sqrt{m_4}$ is the normalized geometric mean curvature. The joint probability density $P(\xi,\hat g)$ of asperities with the summit at normalized elevation $\xi = z/\!\!\sqrt{m_0}$ and the normalized mean geometric curvature $\hat g$ was obtained~\cite{greenwood2006w} to be
\be
  P(\xi,\hat g) = \frac{9}{2\sqrt{2\pi}}\sqrt{\frac{\alpha}{\alpha-1}}\;\hat g^3\; \exp\left[-\frac{\alpha}{2(\alpha-1)}\xi^2 + \frac{3\hat g^2}{2}\right]
  \erfc\left(3\sqrt{\frac{(\alpha-1)}{2(2\alpha-3)}}\left\{ \hat g + \frac{\sqrt\alpha}{3(\alpha-1)}\xi\right\}\right).
  \label{eq:intP}
\ee

From Figs.~\ref{fig:area_04}(b),\ref{fig:area_08}(b) it follows that for a given pressure, the true contact area for a surface with lower value of $\alpha$ is bigger than this for a surface with bigger $\alpha$, i.e. the contact area decreases with increasing value of Nayak parameter. This is also predicted by asperity-based models~\cite{greenwood2006w,carbone2008jmps}. Second, it is clear that the effect of $\alpha$ in asperity based models is exaggerated, and that accurate simulations predict a softer dependence on this parameter. On the other hand, Persson's model, which is independent of $\alpha$, strongly underestimates the true contact area.
As was already reported~\cite{yastrebov2012pre,yastrebov2015ijss} the dispersion of results (see error bars in Figs.\ref{fig:area_04},\ref{fig:area_08}) is higher for smaller $k_l$, smaller $k_s/k_l$ and bigger $H$; for the latter, compare Figs. \ref{fig:area04} and \ref{fig:area08} in Appendix.
In all following figures only corrected contact area is used.

\begin{figure}[htb!]
 \includegraphics[width=1\textwidth]{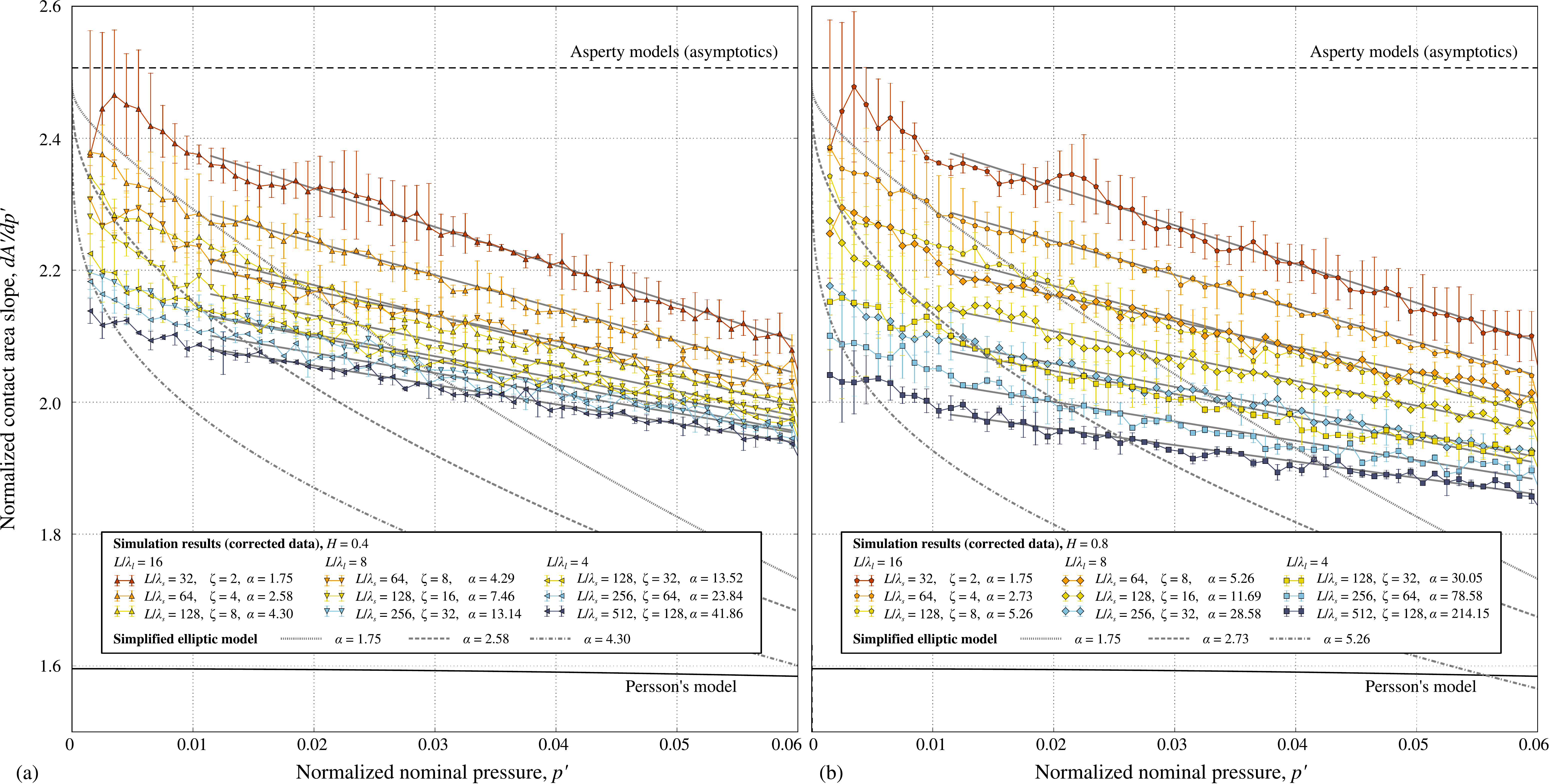}
 \caption{\label{fig:slope}Evolution of the true contact area derivative with respect to the normalized nominal pressure $dA'/dp'$ for (a) $H=0.4$, (b) $H=0.8$. 
 Simulation results (lines with points) are compared with  models: Persson's model (solid line), asymptotic linear solution of asperity models~\cite{bush1975w,carbone2008jmps} (dashed line) and the Greenwood's simplified elliptic model~\cite{greenwood2006w} computed for corresponding Nayak parameters $\alpha$ (gray dashed and dotted lines).
 Grey lines behind simulation points correspond to linear fit in the interval $p'\in[0.01,0.06]$ (see Fig.~\ref{fig:coeffs} and Tables~\ref{tab:1},\ref{tab:2} in Appendix).
 }
\end{figure}

\begin{figure}[htb!]
 \includegraphics[width=1\textwidth]{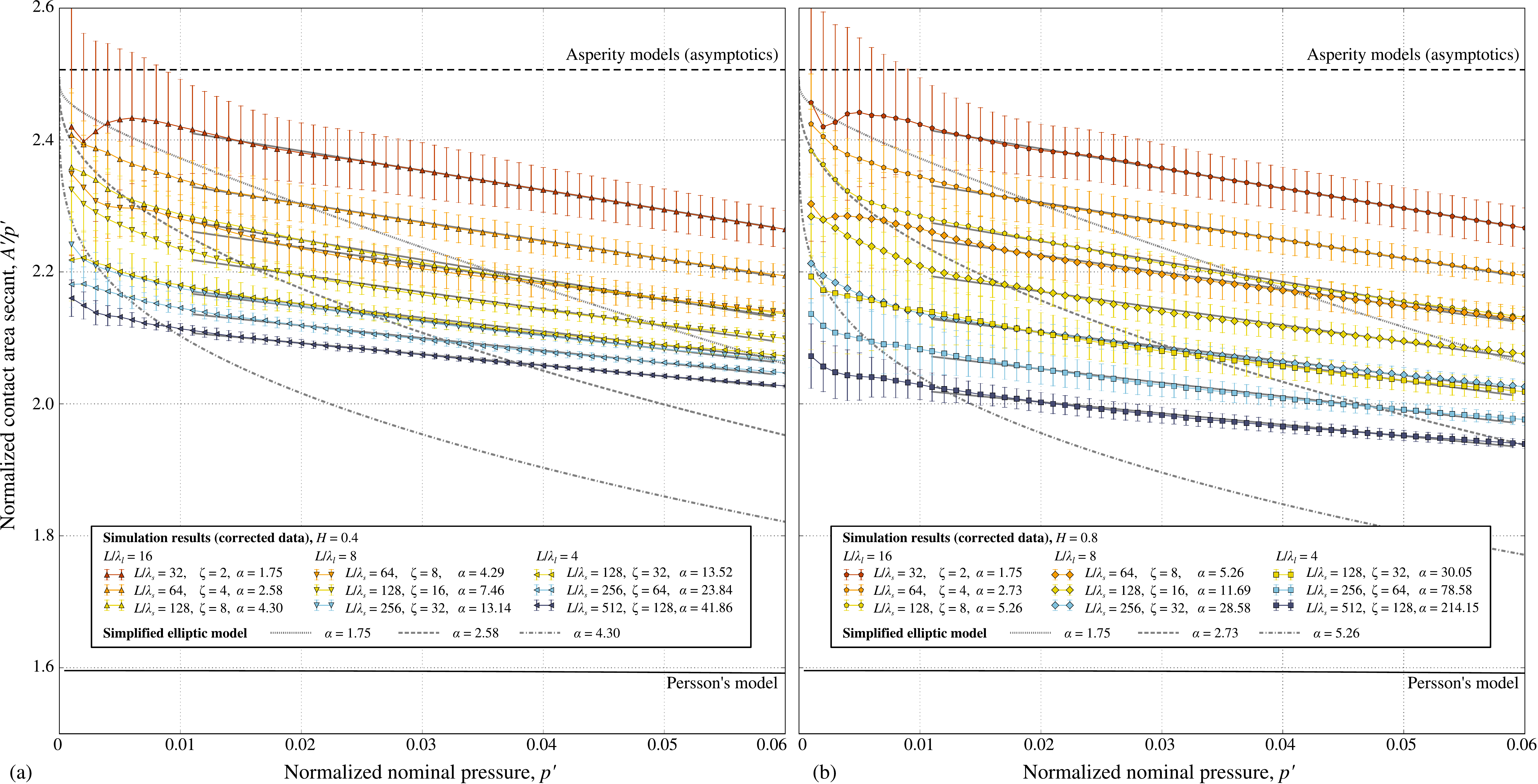}
 \caption{\label{fig:secant}Evolution of the normalized mean inverse pressure $A'/p'$ (contact-area secant) for (a) $H=0.4$, (b) $H=0.8$. 
 Simulation results (lines with points) are compared with analytical models: Persson's model (solid line), asymptotic linear solution of asperity models~\cite{bush1975w,carbone2008jmps} (dashed line) and the Greenwood's simplified elliptic model~\cite{greenwood2006w} computed for corresponding Nayak parameters $\alpha$ (gray dashed and dotted lines).
 Grey lines behind simulation points correspond to linear fit in the interval $p'\in[0.01,0.06]$ (see Fig.~\ref{fig:coeffs} and Tables~\ref{tab:1},\ref{tab:2} in Appendix).
 }
\end{figure}

The normalized contact area derivative with respect to the normalized nominal pressure $dA'/dp'$ is presented in Fig.~\ref{fig:slope} for $H=0.4$ and $H=0.8$. 
The normalized inverse mean contact pressure $A'/p'$ is presented in Fig.~\ref{fig:secant} for the same Hurst exponents. 
These results are compared with i) asymptotic prediction of Nayak's based asperity model: $\lim_{A'\to0}(dA'/dp') = \lim_{A'\to0}(A'/p') = \sqrt{2\pi}$, ii) Persson's model:
\be
  A'_P/p' = \erf(p'\sqrt2)/p',\quad dA'_P/dp' = \sqrt{8/\pi} \exp(-2p'^2)
  \label{eq:persson_model}
\ee
and iii) with the simplified elliptic model~\cite{greenwood2006w}, which was numerically integrated for corresponding Nayak parameters.
As seen from the figures, the contact area slope in the interval $p' \in [0,0.06]$ and $A' \in [0,\approx 13\%]$ decreases by approximately $10-16$ \%, 
i.e. the contact area can be nowhere approximated by an affine function of pressure. 
For a given pressure the slope value decreases with increasing $\alpha$.
Simulation results show higher slopes and softer dependence on Nayak parameter than all asperity-based models, among which the simplified elliptic model has the softest dependence on $\alpha$~\cite{carbone2008jmps}, that was the reason we picked it up for our comparison.
Persson's model predicts very low slope and its decrease with increasing pressure in the interval $p'\in[0,0.06]$ is also considerably smaller than what is numerically obtained.
Results of the inverse normalized mean pressure (the secant of the area-pressure curve) $A'/p'$, which are presented in Fig.~\ref{fig:secant}, demonstrate qualitatively the same results as for the slope, the same conclusions hold.

It could be also remarked that qualitatively the simulation results are consistent with predictions of the simplified elliptic model~\cite{greenwood2006w} computed for small $\alpha$ as was also shown in~\cite{yastrebov2016w}: the slope decreases almost linearly starting from very small pressures, i.e. roughly it behaves as $dA'/dp' \sim a(\alpha) - 2b(\alpha)p'$. 
Hence, by integrating it, the contact area evolution can be approximated by a second order polynomial without zero-order term: $A' = a p' - b p'^2$. 
As seen in Fig.~\ref{fig:secant}, the contact-area secant can be also approximated by an affine function of pressure with approximately the same constants $a'$ and $b'$: 
\be
A'/p' = a' - b'p'. 
\label{eq:secant_linear} 
\ee
In Fig.~\ref{fig:slope},\ref{fig:secant} these linear fits are also plotted.
If the contact area evolution would really be simply the second order polynomial, then $a'=a$ and $b'=b$. Comparison of these coefficients found by the least squares fit in the interval $p'\in[0.006,0.06]$ does demonstrate that $a' \approx a$, but shows a considerable difference in coefficients $b$ and $b'$ (see Fig.~\ref{fig:coeffs} and Tables~\ref{tab:1},\ref{tab:2} in Appendix).
All coefficients can be approximated by a monotonically decreasing logarithmic functions of Nayak parameter, which should be rather universal, thus we give here the identified numerical values (see fit curves in Fig.~\ref{fig:coeffs}):
\be
\begin{split}
  &a'(\alpha) = 2.35 - 0.057\ln(\alpha-1.5),\\
  &b'(\alpha) = 2.85 - 0.24\ln(\alpha-1.5),
\end{split}
  \label{eq:coeff_sec}
\ee
\be
  \begin{split}
  &a(\alpha) = 2.31 - 0.057\ln(\alpha-1.5),\\
  &b(\alpha) = 2.17 - 0.2\ln(\alpha-1.5).
\end{split}
  \label{eq:coeff_slope}
\ee
The coefficients for the contact-area secant~\eqref{eq:coeff_sec} $a',b'$ were identified for $\alpha < 600$, but could be extrapolated up to $\alpha \le 1.5 + \exp(2.85/0.24) \approx 143\,600$ where $b'$ turns zero, the following increase in $\alpha$ would result in a negative $b'$, which is probably nonphysical.
It can be seen in Fig.~\ref{fig:coeffs} that for the linear coefficients $a,a'$ no significant difference between surfaces with different Hurst exponent can be seen,
whereas for the quadratic coefficients $b,b'$ the fit parameters seemingly cluster by the Hurst exponent. An additional study including more surfaces with broader spectrum is needed to draw more accurate conclusions.

Finally, in practice, for a known nominal pressure in elastic contact, the real contact area fraction can be accurately estimated at least in the interval $A'\in[1,15]$ \% by the following phenomenological formula:
\be
  A'(p',m_2,\alpha) = a'(\alpha)p' - b'(\alpha)p'^2,\quad \mbox{with }\quad p' = \frac{p_0}{\sqrt{2m_2}E^*},
\ee
which requires knowing the second spectral moment $m_2$, Nayak parameter $\alpha$ of the effective roughness and the effective elastic modulus of contacting materials.

\subsection{Pressure-dependent coefficient of friction}

Knowing the real contact-area evolution with pressure, one can readily improve an estimation of the static coefficient of friction based on the adhesive theory of friction~\cite{bowden2001b}, in which the true contact interface can bear the maximum shear traction $\sigma_t$ independently of pressure. 
Assuming that at mesoscale a part of the nominally flat frictional interface is subject to pressure $p_0$, the friction coefficient $\mu$ can be defined as the maximal value of the mesoscopic shear traction $t^{\max}$, which the interface can sustain before sliding starts, normalized by the contact pressure $p_0$: 
$$
  \mu(p_0) = \left|\frac{t^{\max}}{p_0}\right| = \frac{A'(p_0)\sigma_t}{p_0} = \frac{a'(\alpha)p' - b'(\alpha)p'^2}{p'E^*\sqrt{2m_2}}\sigma_t
  =  \frac{a'(\alpha)\sigma_t}{E^*\sqrt{2m_2}} - \frac{b'(\alpha)\sigma_t}{2m_2E^*}\cdot\frac{p_0}{E^*},
$$
which can be represented as 
$$
 \mu(p_0) = \mu_0 - \mu_0\frac{b'(\alpha)}{a'(\alpha)\sqrt{2m_2}} \cdot\frac{p_0}{E^*},
$$
where $\mu_0 = a'(\alpha)\sigma_t / E^*\sqrt{2m_2}$ is pressure-independent coefficient of friction and $\mu_0b'(\alpha)/a'(\alpha)\sqrt{2m_2}$ corresponds to the non-linear friction part, which is proportional to $\mu_0$ with a factor depending only on roughness characteristics $m_2$ and $\alpha$. 
Alternatively, using~\eqref{eq:coeff_sec} the pressure-dependent coefficient of friction can be estimated as:
$$
 \mu(p') = \mu_0 \left[1- \beta p'\right].
$$
with $\beta =   [2.85 - 0.24\ln(\alpha-1.5)]\;/[2.35 - 0.057\ln(\alpha-1.5)]$, which can be estimated as follows:
$\beta \approx 0.84$ for $\alpha = 10^2$, $\beta \approx 0.61$ for $\alpha = 10^3$ and $\beta \approx 0.35$ for $\alpha = 10^4$.

\begin{figure}[htb!]
 \includegraphics[width=1\textwidth]{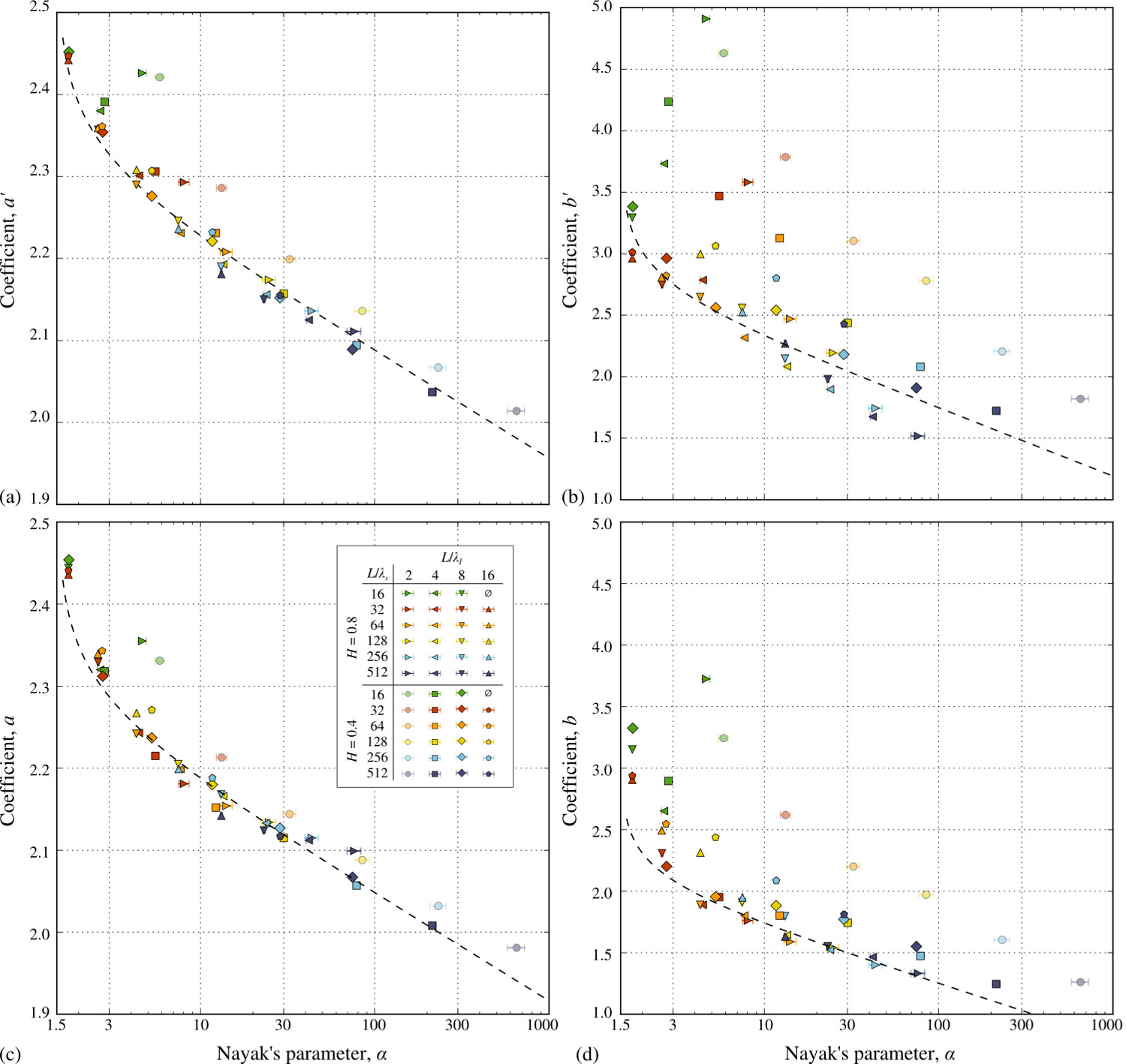}
 \caption{\label{fig:coeffs}(a,b) Coefficients $a',b'$ for the linear fit of the contact area secant $A'/p'$ identified by the least square fit plotted with respect to Nayak parameter $\alpha$ (semi-logarithmic scale) and the fit curve~\eqref{eq:coeff_sec};
 (c,d) coefficients $a,b$ for the linear fit of the contact area slope $dA'/dp'$ identified by the least square fit plotted with respect to Nayak parameter $\alpha$ (semi-logarithmic scale) and the fit curve~\eqref{eq:coeff_slope}. Points of strongly non-representative surfaces with $H=0.8$ and $L/\lambda_l=2$ are made semi-transparent.}
\end{figure}

\subsection{Role of Nayak parameter}

\begin{figure}[htb!]
 \includegraphics[width=1\textwidth]{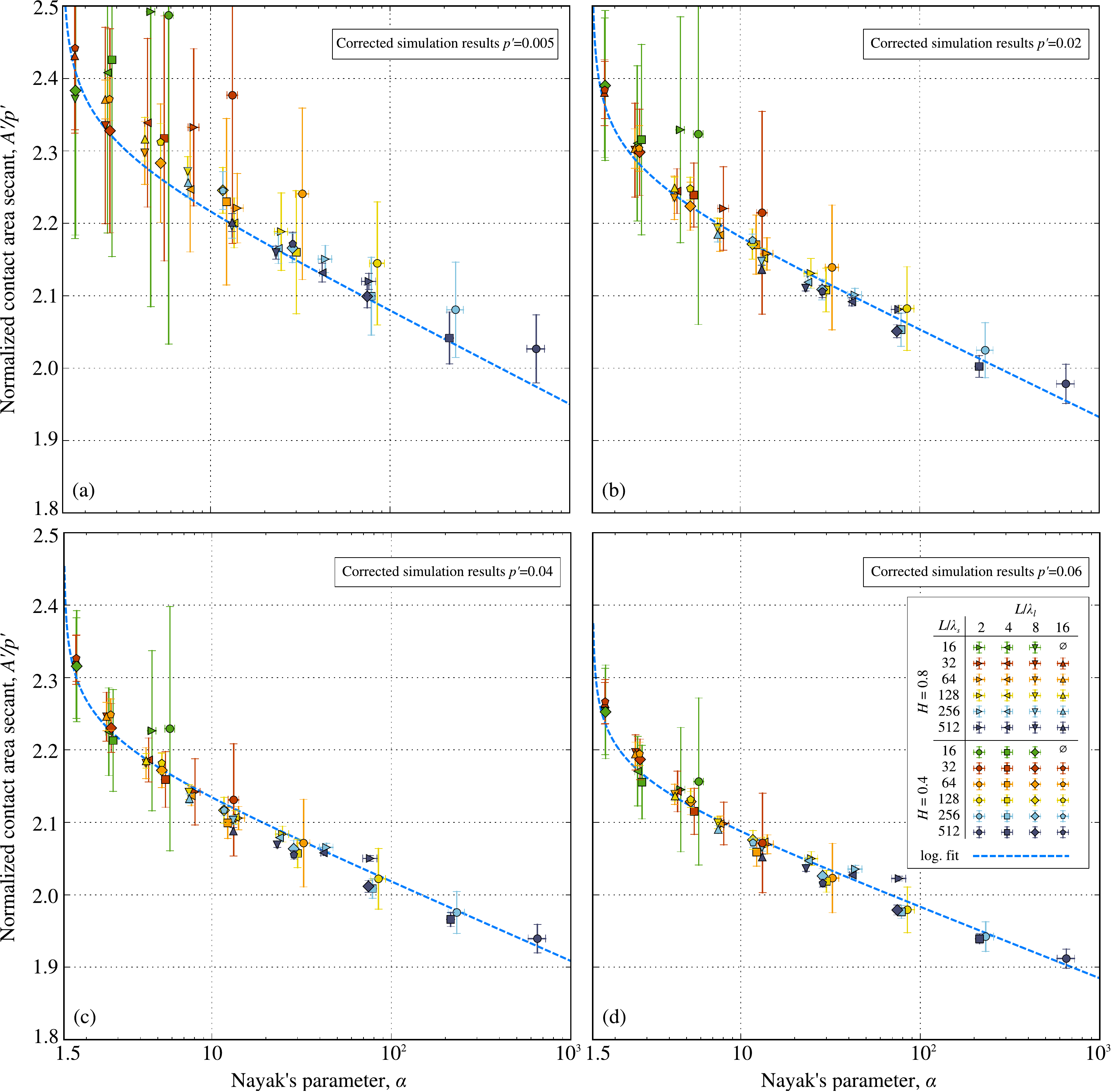}
 \caption{\label{fig:alpha_secant}Secant of the area-pressure curve $A'/p'$ plotted with respect to Nayak parameter $\alpha$ evaluated at (a) $p'=0.005$, (b) $p'=0.02$, (c) $p'=0.04$, (d) $p'=0.06$, mean values and error bars issued from all available simulations are presented; a logarithmic fit (dashed line) corresponds to~\eqref{eq:alpha_fit} with coefficients~\eqref{eq:alpha_coef} found by logarithmic fit of coefficients $a',b'$~\eqref{eq:coeff_sec}.}
\end{figure}

Change of the inverse value of the normalized mean contact pressure $A'/p'$ (the secant of the normalized contact-area/pressure curve) with respect to Nayak parameter $\alpha$ was computed for different normalized nominal pressures $p' = 0.005, 0.02, 0.04, 0.06$ is presented in Fig.~\ref{fig:alpha_secant}(a,b,c,d), respectively. 
These nominal pressure values roughly correspond to contact areas $A' \approx 1, 5, 9, 13$ \%, respectively, computed for a surface with $\alpha \approx 1.75$.
The value of the secant decays with Nayak parameter following for each pressure a logarithmic curve independently of the Hurst exponent, which is well approximated by:
\be
\frac{A'}{p'} = d(p') - c(p')\ln(\alpha-1.5),
\label{eq:alpha_fit}
\ee
with constants $d(p'),c(p')$ which can be expressed through the logarithmic approximations of $a',b'$~\eqref{eq:coeff_sec}: 
\be
d(p') = 2.35 - 2.85p', \quad c(p') = 0.057 - 0.24p'.
\label{eq:alpha_coef}
\ee
Note that points corresponding to $L/\lambda_l = 2,\;L/\lambda_s = 16$, which are not representative and have a rather poor spectrum are located quite far from the master curve.
The remaining points lie very close to~\eqref{eq:alpha_fit} and we expect that the for higher Nayak parameters this logarithmic trend would be preserved.

\begin{figure}[htb!]
 \includegraphics[width=1\textwidth]{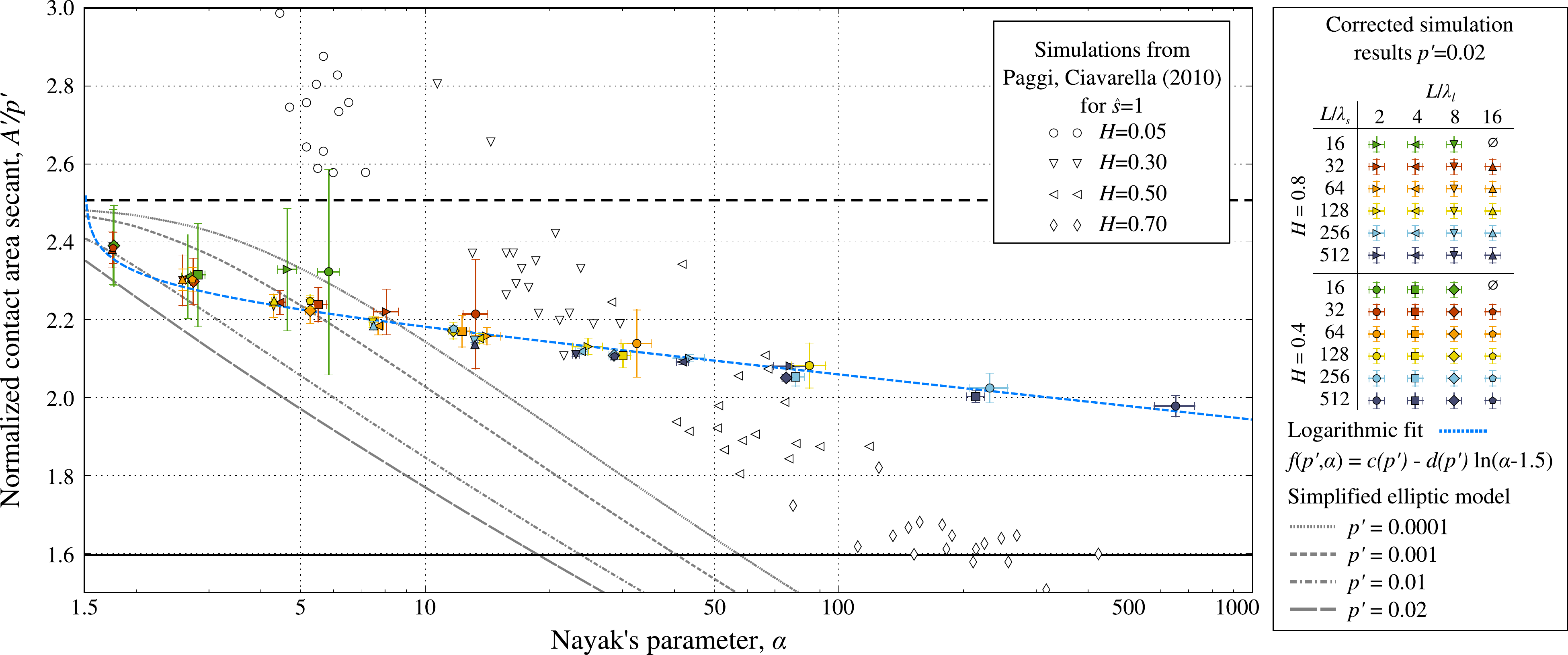}
 \caption{\label{fig:res_sec}Secant of the area-pressure curve $A'/p'$ plotted with respect to Nayak parameter for $p'=0.02$
 mean values and error bars issued from all available simulations are presented as well as a logarithmic fit (dashed line)~\eqref{eq:alpha_fit}.
 Numerical results are compared with results from~\cite{paggi2010w} for separation $\hat s = 1$ and the simplified elliptic model~\cite{greenwood2006w} evaluated at comparable pressures; Persson's model is also computed for the same $p'$.}
\end{figure}

The contact area secant $A'/p'$ evaluated at nominal pressure $p'=0.02$ is compared with numerical results reported in~\cite{paggi2010w} and with predictions of the simplified elliptic model evaluated for comparable pressures (Fig.~\ref{fig:res_sec}). 
Asymptotic solution for asperity based model~\cite{bush1975w,carbone2008jmps} and Persson's solution, both independent of Nayak parameter, are also plotted for reference purpose.
The inverse normalized mean pressure changes very slightly with Nayak parameter $\alpha$, nevertheless this change takes place and is well defined. 
Contrary to what was argued in~\cite{bush1975w}, realistic rough surfaces posses very broad spectra, hence $\alpha$ can be very high~\cite{carbone2008jmps}.

The discrepancy between our results and results obtained in~\cite{paggi2010w} (also presented in Fig.~\ref{fig:res_sec}) comes from the fact that in the latter the rms gradient  was underestimated because it was computed directly on the roughly discretized geometry in the real space:
$$
\sqrt{\langle|\nabla z|^2\rangle} = \frac{1}{N}\sqrt{\sum\limits_{i=0}^{N-1}\sum\limits_{j=0}^{N-1} \left(\frac{z_{i+1,j}-z_{i,j}}{\Delta x}\right)^2 + \left(\frac{z_{i,j+1}-z_{i,j}}{\Delta x}\right)^2},
$$
contrary to the more accurate spectral based computation used here $\sqrt{\langle|\nabla z|^2\rangle} = \sqrt{m_{20}+m_{02}}$, which for a given spectrum does not depend on discretization as far as there are enough modes to represent the spectrum. The same remark concerns the second and the fourth spectral moments needed to determine the Nayak parameter. In addition, no correction of the contact area was used in~\cite{paggi2010w}, and since the authors considered roughnesses with shortest wavelength equivalent to a double spacing between grid points ($\lambda_s = 2\Delta x$), the contact area was overestimated.
In this sense, the results from~\cite{paggi2010w} are rather similar to those reported in~\cite{hyun2004pre}, where a too strong dependence of the secant $A'/p'$ on the Hurst exponent was found (see Fig.~\ref{fig:sec_hurst_compare}). In this light, our results obtained with an accurately evaluated rms gradient and Nayak parameter, and also using the corrective procedure for the contact area computation~\cite{yastrebov2016w} approach better the true solution than the data points extracted from~\cite{paggi2010w}.

\subsection{Effect of Nayak parameter vs Hurst exponent}

\begin{figure}[htb!]
 \includegraphics[width=1\textwidth]{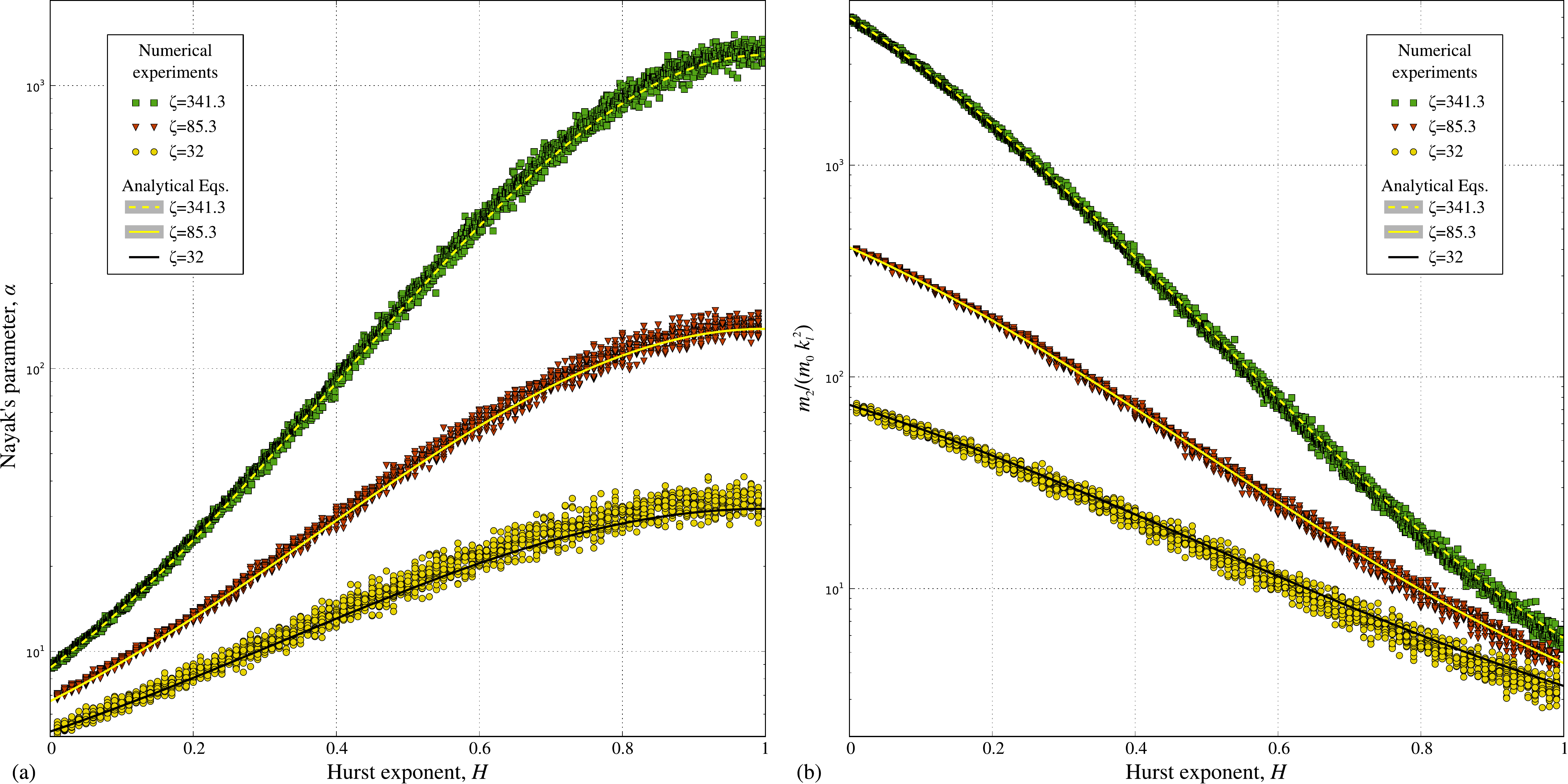}
 \caption{\label{fig:alpha_hurst}Effect of the Hurst exponent (a) on Nayak parameter and (b) on the normalized ratio of spectral moments $m_2/(m_0 k_l^2)$. Points correspond to synthetic rough surfaces (20 realizations for each value of $H$) found for different magnifications $\zeta = 341.3,\;85.3,\;32$, lines correspond to analytical expressions: Eq.~\eqref{eq:a:alpha} for Nayak parameter, and Eq.~\eqref{eq:m2_m0_hurst} for the zeroth and the second spectral moments.}
\end{figure}

In the considered interval of loads the inverse normalized mean pressure (contact area secant, see Fig.~\ref{fig:res_sec}) lies in the interval between Persson's solution and asperity based asymptotic solution as it was already shown by many authors for models of various accuracy and at different pressures~\cite{hyun2004pre,campana2007epl,putignano2012jmps,prodanov2014tl}. However, all these studies report the contact area secant as a function of the Hurst exponent. To the best of our knowledge, there is a unique study~\cite{paggi2010w} in which this secant was plotted with respect to the Nayak parameter (see open points in Fig.~\ref{fig:res_sec}). Hurst exponent of rough surfaces is an independent parameter, but when it changes, the Nayak parameter changes in response. And thus, in all reported studies the change of the secant as a function of the Hurst exponent was probably induced by the change in the Nayak parameter. The link between the Hurst exponent and the Nayak parameter for surfaces without plateau can be expressed as follows~\cite{yastrebov2015ijss}:
\be
  \alpha(H,\zeta) =\frac{3}{2}
  \frac{(1-H)^2}{H(H-2)}\frac{(\zeta^{-2H}-1)(\zeta^{4-2H}-1)}{(\zeta^{2-2H}-1)^2}
\label{eq:a:alpha} 
\ee 
which is a monotonically increasing function of $H$ in the interval $[0,1]$, i.e. higher $H$ corresponds to higher $\alpha$.
In Fig.~\ref{fig:alpha_hurst}(a) the link between the Hurst exponent and the Nayak parameter is shown.
The theoretical prediction~\eqref{eq:a:alpha} is compared with Nayak parameters evaluated through spectral moments on numerous randomly generated rough surfaces with discretization $ 4096 \times 4096 $ and parameters $L/\lambda_l = 6$, $L/\lambda_s = \{192, 512, 2048\}$ corresponding to $\zeta = \{32, 85.33, 341.33\}$ and for changing $H\in[0,1]$; for every value of $H$, 20 surfaces were generated. A good agreement between Eq.~\eqref{eq:a:alpha} and the direct evaluation of the Nayak parameter is observed.
The second and the zero-th spectral moments $m_2$ and $m_0$ take the form~\cite{yastrebov2015ijss}:
\be
  m_2 = \pi \Phi_0  k_l^2 \frac{\zeta^{2-2H}-1}{2(1-H)},\quad m_0 = \pi \Phi_0 \frac{\zeta^{2H}-1}{H\zeta^{2H}},
\label{eq:m2_m0_hurst}
\ee
where $\Phi_0 = \Phi(k_l)$ is the value of PSD calculated at $k_l$.
These estimations are in good agreement with the spectral moments directly evaluated on similarly generated random rough surfaces: the ratio $m_2/(m_0 k_l^2)$ is plotted in Fig.~\ref{fig:alpha_hurst}(b).

\begin{figure}[htb!]
 \includegraphics[width=1\textwidth]{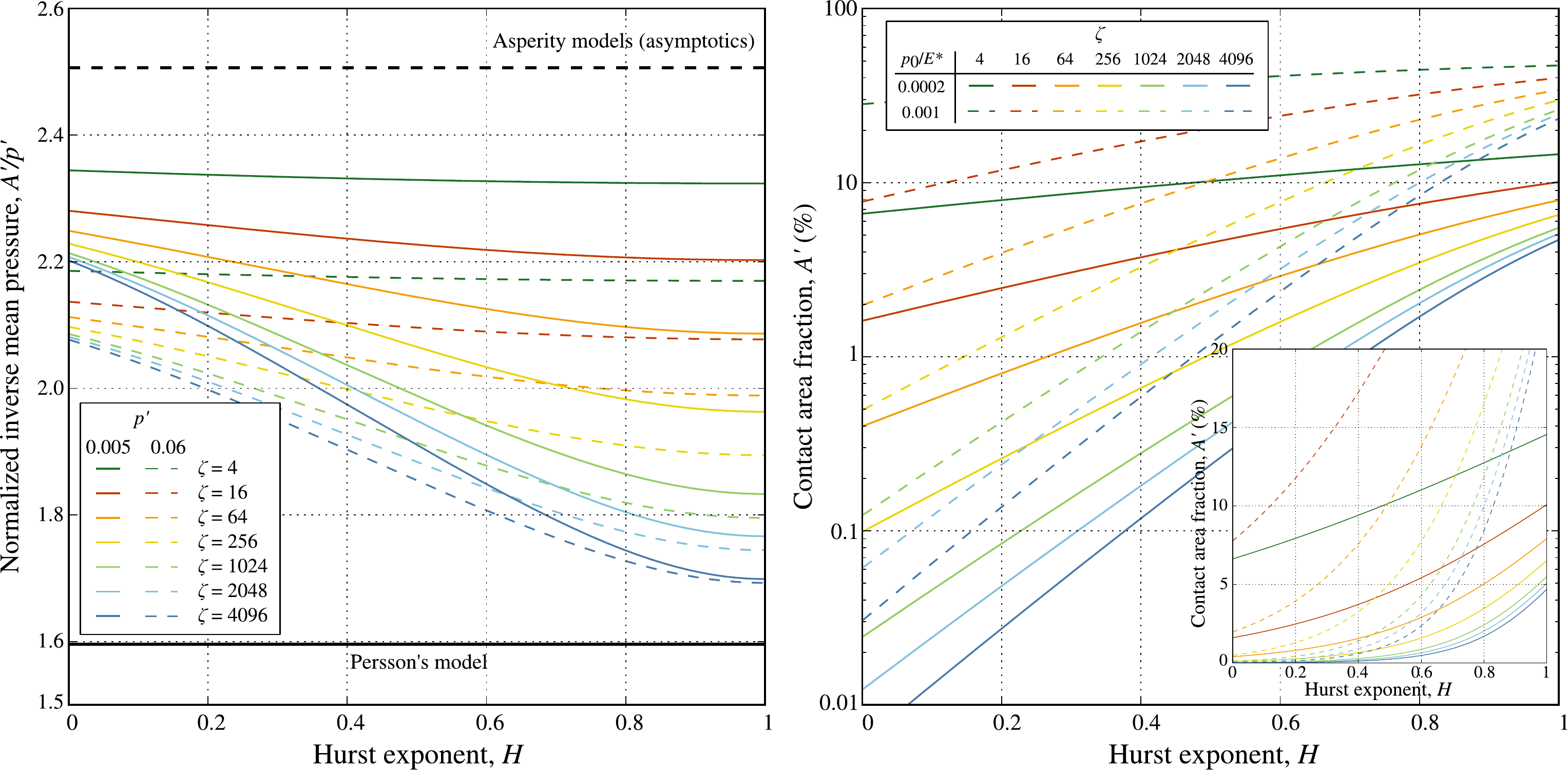}
 \caption{\label{fig:sec_hurst}Effect of the Hurst exponent (a) on the inverse normalized mean pressure $A'/p'$ (the contact-area secant) evaluated for $p'=0.005,0.06$, solid and dashed lines, respectively; and (b) on the true contact area $A'$ (log-scale) for different magnifications $\zeta = k_s/k_l$; the contact area evolution is presented in normal scale in the inset. Equation~\eqref{eq:area_hurst} was used. The contact area is computed for $\Phi_0k_l^2 = 10^{-6}$ and $p_0/E^* = 0.0002, 0.001$ which correspond to solid and dashed lines, respectively.}
\end{figure}

\begin{figure}[htb!]
\begin{center}
 \includegraphics[width=0.5\textwidth]{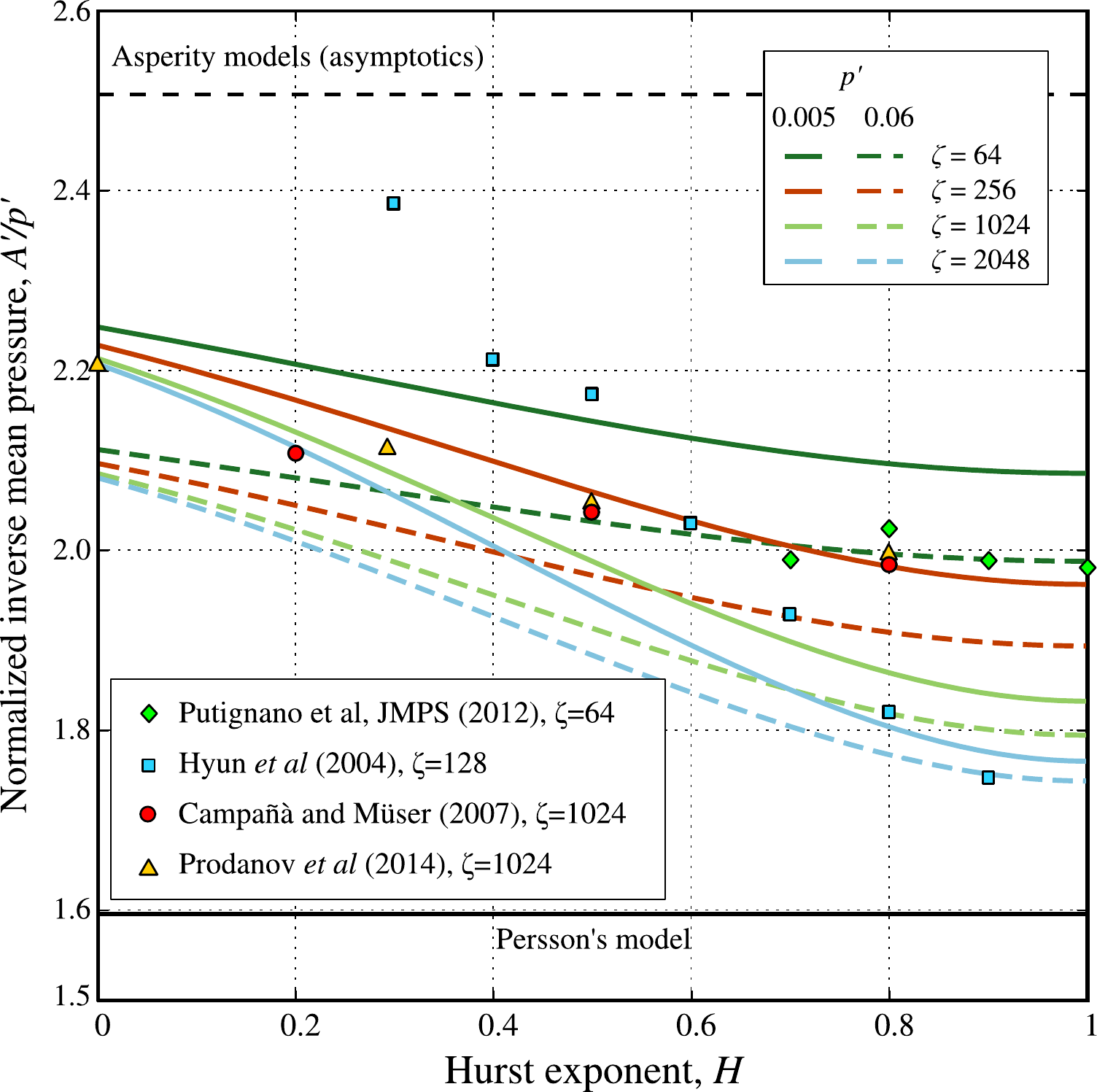}
\end{center}
 \caption{\label{fig:sec_hurst_compare}The contact-area secant $A'/p'$ evaluated for different magnifications $\zeta$ with respect to the Hurst exponent using phenomenological equation~\eqref{eq:area_hurst}, which is based on our numerical results: solid lines correspond to $p'=0.005$, dashed lines correspond to $p'=0.06$. The results are compared with numerical results of other authors (points)~\cite{hyun2004pre,campana2007epl,putignano2012jmps,prodanov2014tl}.
 Asymptotic value of asperity based models and Persson's asymptote are plotted for reference purpose.
 }
\end{figure}

Using Eq.~\eqref{eq:a:alpha}, it can be shown that at extreme values $H=0$ and $H=1$, parameter $\alpha$ takes the following values:
$$
  \alpha(H=0) = -\frac{3}{4} \frac{(\zeta^{4}-1)}{(\zeta^{2}-1)^2} \lim\limits_{H\to0} \frac{(\zeta^{-2H}-1)}{H} = 
  \frac{3}{2}\frac{(\zeta^{2}+1)\ln(\zeta)}{(\zeta^{2}-1)}
$$
$$
  \alpha(H=1) = \frac 3 2(1-\zeta^{-2})(\zeta^{2}-1) \lim\limits_{H\to1} \frac{(1-H)^2}{(\zeta^{2-2H}-1)^2} = 
  \frac 3 8 \left(\frac{\zeta^{2}-1}{\zeta\ln\zeta}\right)^2
$$
Therefore, the change in $\alpha$ over the entire range of allowed Hurst exponents $0\le H\le1$ can be expressed as
$$
  \frac{\alpha(H=1)}{\alpha(H=0)} = 
  \frac14\frac{(\zeta^{2}-1)^3 }{(\zeta^{2}+1)\zeta^2\ln^3(\zeta) } 
   \xrightarrow[\zeta \to \infty]{} \frac{\zeta^2}{4\ln^3(\zeta)}
$$
For small values of $\zeta$ the change in $\alpha$ with $H$ is very small, thus the change in area $A'$ with $H$ is also small.
This can be seen in Fig.~\ref{fig:sec_hurst}.
That was the reason that led authors of~\cite{putignano2012jmps,yastrebov2012pre} to an inaccurate conclusion that contact area is independent (or very weakly dependent) of the Hurst exponent.
See Fig.~\ref{fig:sec_hurst_compare} for comparison of results from~\cite{putignano2012jmps} with our phenomenological results based on numerical simulations.
In reality, the contact area depends on the Hurst exponent through Nayak parameter, but this dependence is very sensitive to the magnification $\zeta$:
for small $\zeta$ the role of $H$ on the contact area is hardly detectable, but for high $\zeta$ the dependence is significant.

In Fig.~\ref{fig:sec_hurst}(a) we plot the inverse normalized mean pressure $A'/p'$ for different values of the Hurst exponent $H\in[0,1]$ and for different magnifications $\zeta = k_s/k_l = \{4,16,64,256,1024,2048,4096\}$ for two normalized nominal pressures $p' = 0.005, 0.06$. For this, we used Eq.~\eqref{eq:alpha_fit},\eqref{eq:alpha_coef} with $\alpha$ expressed through the Hurst exponent via Eq.~\eqref{eq:a:alpha}:
\be
  \frac{A'}{p'}  \approx c(p') - d(p') \ln\left(\frac{3}{2}
  \frac{(1-H)^2}{H(H-2)}\frac{(\zeta^{-2H}-1)(\zeta^{4-2H}-1)}{(\zeta^{2-2H}-1)^2}-\frac32
  \right)
  \label{eq:secant_hurst}
\ee

If $p'$ is expressed through the rms gradient (square root of doubled $m_2$) and, next, through the Hurst exponent and the magnification $\zeta$ using Eq.~\eqref{eq:m2_m0_hurst}:
\be
  p' = \frac{p_0}{E^*\sqrt{\langle|\nabla h|^2\rangle}} = \frac{p_0}{E^*\sqrt{2m_2}} = \frac{p_0}{E^*}\left[\frac{1-H}{\pi \Phi_0k_l^2(\zeta^{2-2H}-1)}\right]^{1/2},
  \label{eq:pprime_m2}
\ee
then the true contact area can be found by substituting $p'$ in form~\eqref{eq:pprime_m2} in~\eqref{eq:secant_hurst}:
\be
  A'(p_0/E^*,H,\zeta,\Phi_0k_l^2) \approx a_0(H,\Phi_0k_l^2,\zeta)\frac{p_0}{E^*} - b_0(H,\Phi_0k_l^2,\zeta)\frac{p^2_0}{E^{*2}},
  \label{eq:area_hurst}
\ee
where the coefficients can be found using~\eqref{eq:alpha_coef}:
$$
a_0(H,\Phi_0k_l^2,\zeta) = \left[2.35 - 0.057 \ln\left(\frac{3}{2}
  \frac{(1-H)^2}{H(H-2)}\frac{(\zeta^{-2H}-1)(\zeta^{4-2H}-1)}{(\zeta^{2-2H}-1)^2}-\frac32
  \right)\right] \left[\frac{1-H}{\pi \Phi_0k_l^2(\zeta^{2-2H}-1)}\right]^{1/2},
$$
$$
b_0(H,\Phi_0k_l^2,\zeta) = \left[2.85 - 0.24 \ln\left(\frac{3}{2}
  \frac{(1-H)^2}{H(H-2)}\frac{(\zeta^{-2H}-1)(\zeta^{4-2H}-1)}{(\zeta^{2-2H}-1)^2}-\frac32
  \right)\right]\left[\frac{1-H}{\pi \Phi_0k_l^2(\zeta^{2-2H}-1)}\right].
$$
The true contact areas computed for different pressures $p_0/E^* = 0.0002, 0.001$ and fixed $\Phi_0k_l^2 = 10^{-6}$ are plotted in Fig.~\ref{fig:sec_hurst}(b) as functions of the Hurst exponent for different magnifications $\zeta$.

When reformulated in terms of the Hurst exponent and magnification, our phenomenological results based on very accurate and statistically sound numerical simulations can be compared with similar results of other authors (Fig.~\ref{fig:sec_hurst_compare}). Our data is plotted as bounds on the secant evaluated at nominal pressures $p'=0.005,0.06$ and computed for surfaces with different magnifications $\zeta$. In one of the pioneering works in numerical analysis of rough contact~\cite[Fig. 7]{hyun2004pre} it was shown that the contact-area secant $A'/p'$ (and thus the contact area) evaluated at certain nominal pressure, decreases with increasing Hurst exponent. It was shown that for $H=0.3$ it approached closely the asymptotic limit of asperity models and for $H=0.9$ it was just above Persson's asymptotic value. Other authors reproduced this study and plotted the secant as a function of the Hurst exponent~\cite{campana2007epl,pohrt2012prl,putignano2012jmps,prodanov2014tl}. As already mentioned, in some studies~\cite{putignano2012jmps,yastrebov2012pre}, however, it was argued that for properly discretized surfaces the secant does not depend (or depends very weakly) on the Hurst exponent.
As seen from our results, indeed the dependence on the Hurst exponent is the weakest for $H$ approaching unity. In this light, the numerical results from~\cite{putignano2012jmps} show a reasonable agreement with our predictions. Results from~\cite{campana2007epl,prodanov2014tl} seem to overestimate the contact area as no correction technique was used, which is crucial for accurate numerical simulations, especially for the high magnifications $\zeta = 1024$ used. Finally, the results from~\cite{hyun2004pre} suffer both from inaccurate estimation of rms gradient, too coarse mesh and the absence of area correction.

\subsection{From infinitesimal to full contact}

\begin{figure}[htb!]
 \includegraphics[width=1\textwidth]{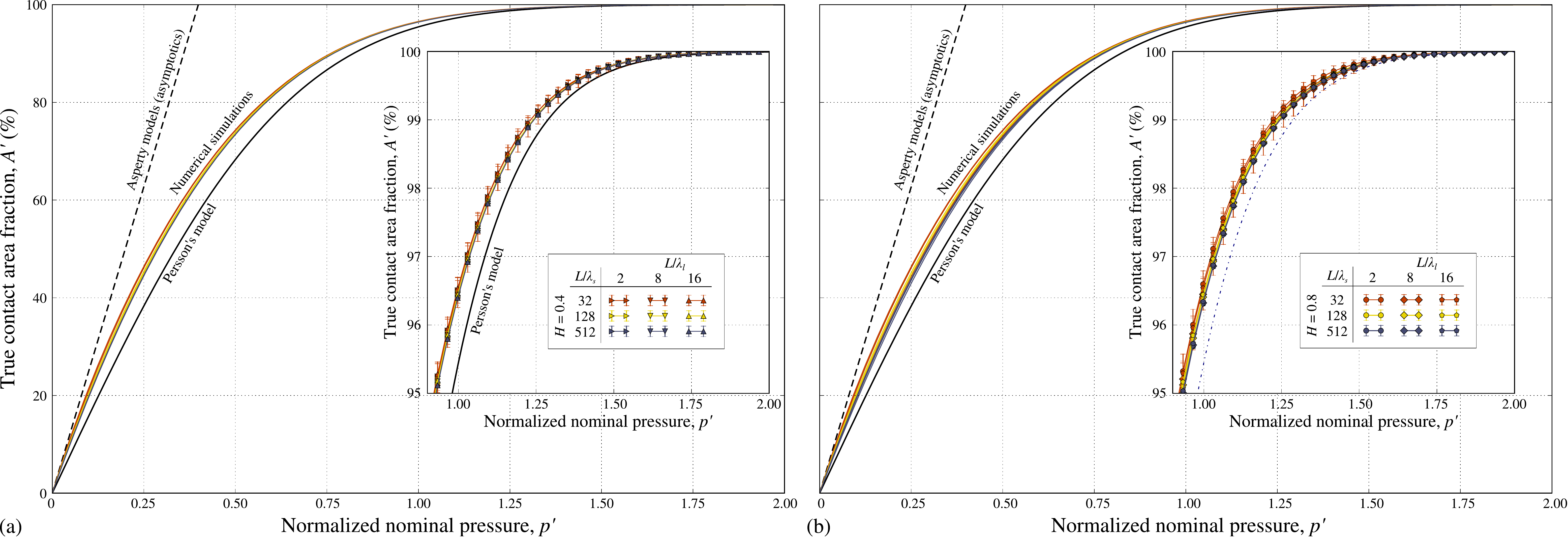}
 \caption{\label{fig:area_full}Contact area evolution from infinitesimal to full contact compared with the prediction of Persson's model: (a) for $H=0.4$, (b) for $H=0.8$. Asymptotic solution of asperity models is also plotted for reference purpose. Zoomed area near the full contact is shown in the insets.}
\end{figure}

Contact area evolution up to full contact is depicted in Fig.~\ref{fig:area_full} and compared with Persson's prediction~\eqref{eq:persson_model}.
Qualitatively Persson's model describes properly the evolution of the true contact area but quantitatively, simulations predict higher values of area and thus full contact occurs earlier in simulations~\footnote{Since Persson's model deals with a continuous spectrum and thus with a truly Gaussian surface with an infinite support, full contact happens only at infinite nominal pressure, but since error function tends very rapidly to unity, a certain threshold can be used to identify full contact, i.e. it could be $A'_{\mbox{\tiny th}} = 1-1/N^2$, which corresponds to the last element coming in contact, where $N$ is the number of grid points per side.}. Near full contact, shown in the inset in Fig.~\ref{fig:area_full}, all results collapse in a single master curve, which is well described by the queue of an error function. 

The results near full contact can be of interest in metal forming applications~\cite{bay1987friction,wilson1988real,azushima2016tribology} but also in theoretical findings~\cite{greenwood2015almost,xu2017statistical,ciavarella2016rough}.
However, near full contact our results have to be handled with precaution, because there, the area correction technique~\cite{yastrebov2016w} might work less properly. 
Due to insufficient discretization if a non-contact patch, surrounded by a contact zone, decreases in size under increasing load, it will soon disappear when reached the diameter of one-two elements (in raw simulations the contact area is overestimated). When such a patch disappears, the associated contact area cannot be corrected because there is no perimeter associated with this patch.

\subsection{Implications for Persson's model}

\begin{figure}[htb!]
 \includegraphics[width=1\textwidth]{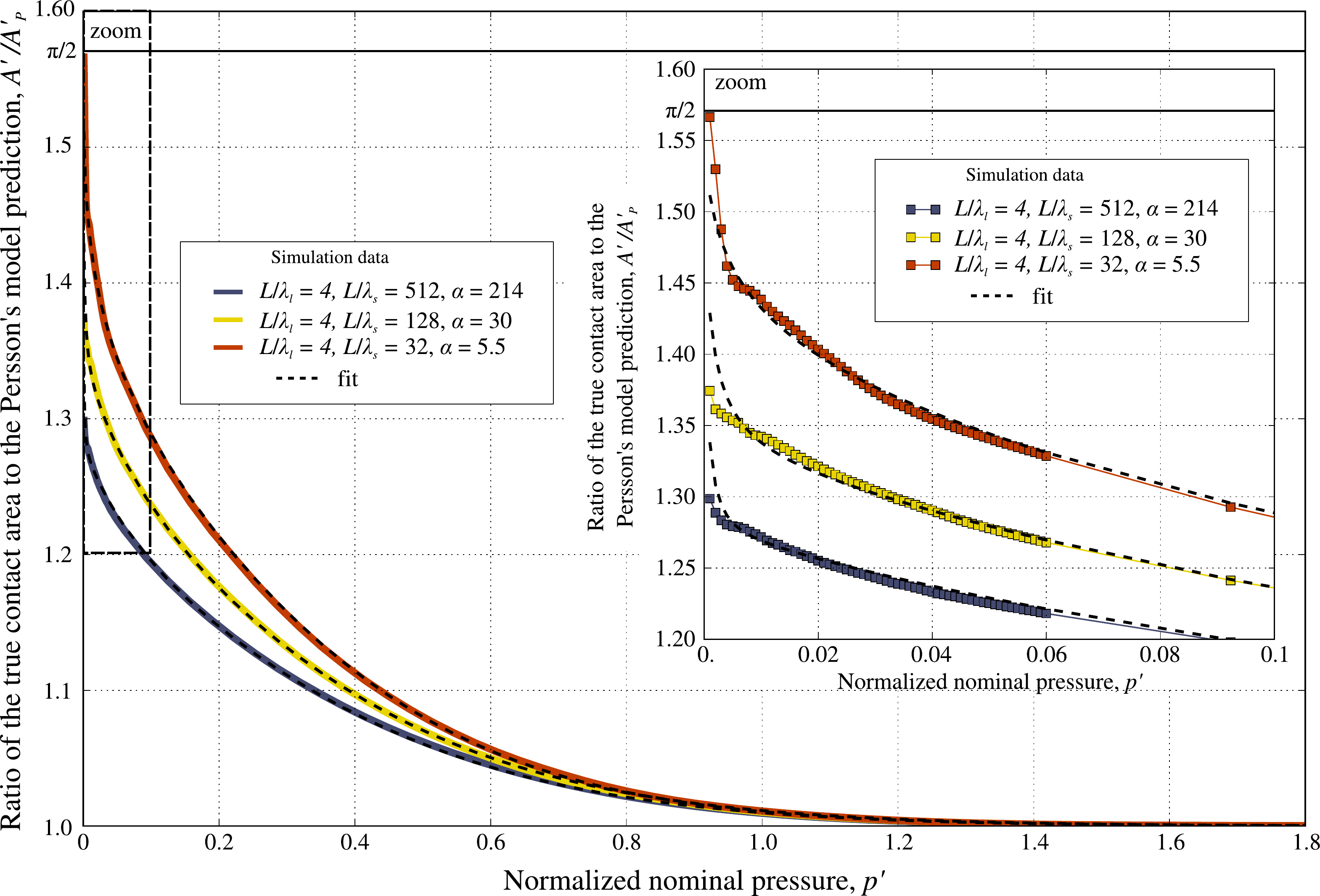}
 \caption{\label{fig:area_persson}Ratio between the contact area computed numerically with unprecedented accuracy and Persson's model for different surface spectra (bold colored curves), dashed lines represent the fitted Eq.~\eqref{eq:gamma_persson}.}
\end{figure}

Persson's model, derived for full contact under infinite nominal pressure, deals with the probability density of contact pressures $P(p)$, which
takes the forms of a diffusion equation, where $P$ acts as the concentration of diffusing quantity and the variance of the surface gradient $m_2$ acts as time.
In a simplified form obtained in~\cite{manners2006w}, it writes as:
$$
  \ddp{P(p,m_2)}{m_2} = \frac{E^{*2}}{4}\ddp{^2P(p,m_2)}{p^2}
$$
This interesting result, was extended to partial contacts and thus finite pressures, by imposing the probability density of zero pressure to be zero, i.e. $P(0,m_2) = 0$.
This extension, however, is the weakest point of the model as this transition from full to partial contact is not justified~\cite{manners2006w}. 
Numerous comparisons of Persson's model with numerical simulations~\cite{hyun2004pre,pei2005jmps,campana2007epl,prodanov2014tl,campana2008jpcm,putignano2012jmps,yastrebov2012pre,yastrebov2015ijss}, especially see~\cite{dapp2014jpcm}, revealed that the contact area predicted by Persson's model is underestimated. Meanwhile, Persson argued that introducing the boundary condition is not sufficient for extending his model to partial contacts, but that it is also needed to take into account the interfacial elastic energy change due to detachment~\cite{persson2006contact,yang2008jpcm}. 
It was argued that the ``scaling factor'' needed to adjust the energy is of ``order of unity''. 
In Fig.~\ref{fig:area_persson} we plot the ratio of the numerically predicted contact area to the prediction of the original Persson's model. 
Three surfaces are considered $L/\lambda_l = 4$, $H=0.8$ and $L/\lambda_s = \{32, 128, 512\}$ corresponding to $\alpha \approx \{5.5, 30, 214\}$.
This ratio evolves non-linearly with increasing pressure and at the same time it strongly depends on Nayak parameter.

An approximate form of this ratio takes the following form:
\be
f(p') = 1+\left([c_1+(1-c_1)\exp(-c_2\sqrt{p'})]\frac{\pi}{2}-1\right)\exp(-c_3p')(1+c_4p'^{3/2}),
\label{eq:gamma_persson}
\ee
and could be used as an alternative phenomenological correction~\cite{persson2006contact,persson2016multiscale}.
Note that as $p'\to\infty$ then $f(p') \to 1$, and as $p'\to0$ then  $f(p') \to \pi/2$, i.e. $f(p')$ tends to the ratio between asperity models' asymptote to the Persson's asymptote.
Eq.~\eqref{eq:gamma_persson} includes two exponential decays: the first one, decaying with the square root of pressure, is relevant to the interval of small pressure, especially for surfaces with high Nayak parameters, the second decay partly balanced by a power-law with an exponent $3/2$ is responsible for the long range decay up to full contact. This non-trivial two-stage correction demonstrates that the ``factor of order of unity'' is not simply a factor but a complex function depending on Nayak parameter, whose introduction would be hard to justify.
In Fig.~\ref{fig:area_persson} the following coefficients were used to fit the ratio of the simulated true contact area to the Persson's prediction:
\begin{itemize}
 \item $L/\lambda_s = 32$,  $\alpha\approx5.5$: $c_1=0.905$, $c_2=15$, $c_3 = 5.7$, $c_4=6.5$;
 \item $L/\lambda_s = 128$, $\alpha\approx30$: $c_1=0.855$, $c_2=30$, $c_3 = 5.2$, $c_4=5.0$;
 \item $L/\lambda_s = 512$, $\alpha\approx214$: $c_1=0.815$, $c_2=50$, $c_3 = 5.2$, $c_4=5.4$.
\end{itemize}
all three roughnesses have $L/\lambda_l = 4$ and $H=0.8$.

\section{Conclusion\label{sec:discusion}}

By carrying out statistically sound high accuracy numerical simulations of mechanical contact between rough surfaces with determined spectral content, we demonstrated how Nayak parameter $\alpha$ affects the true contact area evolution with increasing nominal pressure. 
We showed that in the interval $\alpha \in [1.75,\;\approx600]$, an increase in $\alpha$ results in a decrease in the contact area fraction $A'$. 
This trend was already predicted by advanced asperity models based on Nayak's random process model, but due to inherent approximations of asperity models, this dependence was exaggerated. On the contrary, the original Persson's model of rough contact does not take consideration of Nayak parameter. 
We deduced that the contact area for a given normalized pressure decreases logarithmically with Nayak parameter.

In several previous studies of the rough contact, attempts ware undertaken to deduce the role of the Hurst exponent (or fractal dimension) in the true contact area. 
In some studies it was argued that the contact area might be independent or too weakly dependent on the Hurst exponent. 
Here we show clearly the link between the Nayak parameter and the Hurst exponent, and also we demonstrate that for surfaces with high ``magnification'', i.e. surfaces containing many modes $\zeta = k_s/k_l \gg 1$, Nayak parameter changes drastically with the Hurst exponent. Thus, the latter has a strong impact on the contact area for surfaces with a rich spectrum, but this impact, we believe, is determined mainly by the corresponding change in Nayak parameter, and not by the Hurst exponent itself. 
A study of the rough contact for higher Nayak parameters, which possibly can discriminate independent effects of Nayak parameter and Hurst exponent, is still missing.

As a by-product of this study, we suggested a phenomenological second order polynomial estimation of the contact area evolution, which is valid at least in the interval $A'\in[1,15]$ \%.
The coefficients of this equation decay logarithmically with the Nayak parameter, the associated universal constants are also determined.
Based on the contact area shape, we gave an explicit formula for a pressure-dependent coefficient of friction.
Finally, we showed that the correction of Persson's model for partial contact in a ``phenomenological way'' is not an easy task, because of the lack of dependence on Nayak parameter and because of a complex nonlinear evolution of the true contact area, which cannot be accurately described by the error function. 

\vspace{1cm}
\noindent{\bf \Large Acknowledgment}\\[5pt]\normalsize

This paper was partly presented at 24$^{\mbox{\tiny th}}$  International Congress of Theoretical and Applied Mechanics (ICTAM) which was hold in Montr\'eal, Canada, and VY is grateful to Association Fran\c caise de M\'ecanique (AFM), Comit\'e National Fran\c cais de M\'ecanique (CNFM) and to CNRS for the travel support.

\vspace{1cm}
\appendix
\noindent{\bf \Large Appendix}\\[5pt]\normalsize

\section{Numerical data}

In Tables~\ref{tab:1},\ref{tab:2}, the coefficients for an affine fit of the inverse mean pressure $A'/p'$ evolution (up to $\approx 15$ \% of contact area) are listed for $H=0.4$ and $H=0.8$, respectively. The coefficients are found by the mean least squared error method in the interval $p'\in[0.006,0.06]$.  These coefficients, depicted in Fig.~\ref{fig:coeffs} are used in Eq.~\eqref{eq:secant_linear} and \eqref{eq:coeff_sec} to define the contact area dependence on Nayak parameter.

\begin{table}[ht!]
 \begin{tabular}{ccccccc}
 $L/\lambda_l$ & $L/\lambda_s$ & $\alpha\pm \mathrm{std}(\alpha)$ & $b$ & $a$ & $b'$ & $a'$ \\
 \hline\\[-10pt]
2 & 16 & 4.64 $\pm$ 0.24 & 3.725 & 2.355 & 4.909 & 2.426\\
2 & 32 & 8.05 $\pm$ 0.55 & 1.761 & 2.181 & 3.581 & 2.293\\
2 & 64 & 14.08 $\pm$ 1.12 & 1.589 & 2.154 & 2.47 & 2.208\\
2 & 128 & 24.75 $\pm$ 2.08 & 1.541 & 2.134 & 2.194 & 2.174\\
2 & 256 & 43.52 $\pm$ 3.81 & 1.4 & 2.115 & 1.744 & 2.136\\
2 & 512 & 76.24 $\pm$ 6.83 & 1.332 & 2.099 & 1.518 & 2.111\\
4 & 16 & 2.66 $\pm$ 0.09 & 2.652 & 2.32 & 3.732 & 2.38\\
4 & 32 & 4.44 $\pm$ 0.14 & 1.889 & 2.243 & 2.786 & 2.301\\
4 & 64 & 7.69 $\pm$ 0.22 & 1.802 & 2.199 & 2.317 & 2.231\\
4 & 128 & 13.52 $\pm$ 0.43 & 1.644 & 2.166 & 2.083 & 2.193\\
4 & 256 & 23.84 $\pm$ 0.79 & 1.523 & 2.133 & 1.897 & 2.156\\
4 & 512 & 41.86 $\pm$ 1.41 & 1.465 & 2.112 & 1.675 & 2.125\\
8 & 16 & 1.75 $\pm$ 0.01 & 3.152 & 2.444 & 3.294 & 2.446\\
8 & 32 & 2.59 $\pm$ 0.04 & 2.308 & 2.329 & 2.749 & 2.357\\
8 & 64 & 4.29 $\pm$ 0.08 & 1.89 & 2.242 & 2.648 & 2.29\\
8 & 128 & 7.46 $\pm$ 0.13 & 1.909 & 2.205 & 2.559 & 2.246\\
8 & 256 & 13.14 $\pm$ 0.22 & 1.797 & 2.168 & 2.148 & 2.19\\
8 & 512 & 23.13 $\pm$ 0.4 & 1.553 & 2.124 & 1.98 & 2.15\\
16 & 32 & 1.75 $\pm$ 0.01 & 2.905 & 2.436 & 2.962 & 2.442\\
16 & 64 & 2.58 $\pm$ 0.02 & 2.494 & 2.339 & 2.808 & 2.359\\
16 & 128 & 4.3 $\pm$ 0.05 & 2.314 & 2.267 & 2.997 & 2.308\\
16 & 256 & 7.49 $\pm$ 0.1 & 1.949 & 2.199 & 2.525 & 2.236\\
16 & 512 & 13.18 $\pm$ 0.2 & 1.63 & 2.142 & 2.27 & 2.181\\
 \end{tabular}
 \caption{\label{tab:1}Coefficients $a,b,a',b'$ for the linear interpolation for the contact area slope $dA'/dp' \approx a - 2bp'$ and for the contact area secant $A'/p' \approx a' - b'p'$ found by the least square fit of numerical results in the interval $p'\in[0.006,0.06]$ identified for $H=0.4$.}
\end{table}

\begin{table}[ht!]
 \begin{tabular}{ccccccc}
 $L/\lambda_l$ & $L/\lambda_s$ & $\alpha\pm \mathrm{std}(\alpha)$ & $b$ & $a$ & $b'$ & $a'$ \\
 \hline\\[-10pt]
2 & 16 & 5.84 $\pm$ 0.35 & 3.244 & 2.331 & 4.63 & 2.421\\
2 & 32 & 13.23 $\pm$ 0.89 & 2.619 & 2.213 & 3.786 & 2.286\\
2 & 64 & 32.45 $\pm$ 2.72 & 2.201 & 2.144 & 3.104 & 2.199\\
2 & 128 & 84.83 $\pm$ 8.0 & 1.971 & 2.088 & 2.78 & 2.136\\
2 & 256 & 231.58 $\pm$ 24.32 & 1.605 & 2.032 & 2.205 & 2.067\\
2 & 512 & 651.48 $\pm$ 73.05 & 1.261 & 1.981 & 1.82 & 2.014\\
4 & 16 & 2.82 $\pm$ 0.11 & 2.896 & 2.318 & 4.236 & 2.391\\
4 & 32 & 5.51 $\pm$ 0.26 & 1.952 & 2.215 & 3.468 & 2.306\\
4 & 64 & 12.26 $\pm$ 0.55 & 1.802 & 2.152 & 3.126 & 2.231\\
4 & 128 & 30.05 $\pm$ 1.33 & 1.744 & 2.115 & 2.437 & 2.157\\
4 & 256 & 78.58 $\pm$ 3.82 & 1.474 & 2.057 & 2.081 & 2.094\\
4 & 512 & 214.15 $\pm$ 10.83 & 1.245 & 2.008 & 1.723 & 2.037\\
8 & 16 & 1.76 $\pm$ 0.01 & 3.325 & 2.454 & 3.383 & 2.452\\
8 & 32 & 2.75 $\pm$ 0.04 & 2.202 & 2.312 & 2.963 & 2.354\\
8 & 64 & 5.26 $\pm$ 0.13 & 1.955 & 2.237 & 2.561 & 2.276\\
8 & 128 & 11.69 $\pm$ 0.32 & 1.883 & 2.18 & 2.541 & 2.221\\
8 & 256 & 28.58 $\pm$ 0.77 & 1.772 & 2.127 & 2.181 & 2.152\\
8 & 512 & 74.54 $\pm$ 2.03 & 1.551 & 2.067 & 1.909 & 2.089\\
16 & 32 & 1.75 $\pm$ 0.01 & 2.938 & 2.441 & 3.012 & 2.447\\
16 & 64 & 2.73 $\pm$ 0.02 & 2.546 & 2.343 & 2.821 & 2.361\\
16 & 128 & 5.26 $\pm$ 0.07 & 2.437 & 2.271 & 3.063 & 2.307\\
16 & 256 & 11.71 $\pm$ 0.18 & 2.086 & 2.188 & 2.801 & 2.232\\
 \end{tabular}
 \caption{\label{tab:2}Coefficients $a,b,a',b'$ for the linear interpolation for the contact area slope $dA'/dp' \approx a - 2bp'$ and for the contact area secant $A'/p' \approx a' - b'p'$ found by the least square fit of numerical results in the interval $p'\in[0.006,0.06]$ identified for $H=0.8$.}
\end{table}

In Tables~\ref{tab:3},\ref{tab:4} we list some numerical results obtained for different rough surfaces for $H=0.4$ and $H=0.8$, respectively. Only every third point is shown, and the data is limited to the interval $p' < 0.06$. The entire set of numerical data is provided in the supplementary material~\cite{supplemental}.

\begin{table}[ht!]
\begin{center}
 \begin{tabular}{ccccccccc}
	 & \multicolumn{2}{c}{$L/\lambda_s=4,L/\lambda_l=32$} 
	 & \multicolumn{2}{c}{$L/\lambda_s=8,L/\lambda_l=128$} 
	 & \multicolumn{2}{c}{$L/\lambda_s=16,L/\lambda_l=512$} 
	 & \multicolumn{2}{c}{$L/\lambda_s=4,L/\lambda_l=512$}\\
	 & \multicolumn{2}{c}{$\alpha = 4.45 \pm 0.15$} 
	 & \multicolumn{2}{c}{$\alpha = 7.46 \pm 0.14$} 
	 & \multicolumn{2}{c}{$\alpha = 13.19 \pm 0.20$} 
	 & \multicolumn{2}{c}{$\alpha = 41.87 \pm 1.41$}\\[3pt]
$p'$    & $A'$  & std$(A')$     & $A'$  & std$(A')$     & $A'$  & std$(A')$     & $A'$  & std$(A')$\\[2pt]\hline\\[-8pt]
0.001	 & 0.245 & 0.031	 & 0.232 & 0.007	 & 0.224 & 0.002	 & 0.216 & 0.003\\
0.004	 & 0.942 & 0.053	 & 0.912 & 0.009	 & 0.883 & 0.006	 & 0.854 & 0.006\\
0.007	 & 1.628 & 0.066	 & 1.579 & 0.011	 & 1.530 & 0.009	 & 1.486 & 0.006\\
0.010	 & 2.299 & 0.076	 & 2.234 & 0.015	 & 2.172 & 0.009	 & 2.114 & 0.008\\
0.013	 & 2.962 & 0.075	 & 2.887 & 0.018	 & 2.806 & 0.011	 & 2.738 & 0.011\\
0.016	 & 3.621 & 0.070	 & 3.533 & 0.021	 & 3.437 & 0.013	 & 3.360 & 0.012\\
0.019	 & 4.273 & 0.061	 & 4.175 & 0.026	 & 4.065 & 0.012	 & 3.978 & 0.012\\
0.022	 & 4.919 & 0.067	 & 4.812 & 0.030	 & 4.686 & 0.012	 & 4.593 & 0.013\\
0.025	 & 5.563 & 0.073	 & 5.447 & 0.032	 & 5.304 & 0.013	 & 5.209 & 0.013\\
0.028	 & 6.203 & 0.079	 & 6.079 & 0.034	 & 5.918 & 0.014	 & 5.818 & 0.012\\
0.031	 & 6.840 & 0.090	 & 6.704 & 0.036	 & 6.531 & 0.015	 & 6.425 & 0.014\\
0.034	 & 7.475 & 0.102	 & 7.328 & 0.040	 & 7.142 & 0.015	 & 7.029 & 0.014\\
0.037	 & 8.110 & 0.113	 & 7.950 & 0.042	 & 7.747 & 0.017	 & 7.631 & 0.014\\
0.040	 & 8.745 & 0.121	 & 8.567 & 0.042	 & 8.352 & 0.019	 & 8.231 & 0.014\\
0.043	 & 9.377 & 0.131	 & 9.180 & 0.043	 & 8.952 & 0.019	 & 8.828 & 0.015\\
0.046	 & 10.003 & 0.137	 & 9.789 & 0.043	 & 9.551 & 0.019	 & 9.423 & 0.015\\
0.049	 & 10.621 & 0.145	 & 10.396 & 0.044	 & 10.148 & 0.021	 & 10.014 & 0.015\\
0.052	 & 11.235 & 0.152	 & 11.000 & 0.045	 & 10.743 & 0.021	 & 10.604 & 0.016\\
0.055	 & 11.845 & 0.159	 & 11.601 & 0.047	 & 11.333 & 0.022	 & 11.192 & 0.016\\
0.058	 & 12.453 & 0.165	 & 12.198 & 0.048	 & 11.921 & 0.024	 & 11.775 & 0.018\\
 \end{tabular}
 \end{center}
 \caption{\label{tab:3}Sampled numerical results for the contact area fraction $A'$ growth with the normalized nominal pressure $p' = p_0/(\sqrt{2m_2}E^*)$ for surfaces with $H=0.4$ and different cutoffs in the spectrum, mean results over 10 realizations is shown as well as the standard deviation denoted std$(A')$. The contact area is corrected using~\eqref{eq:cor_area}.}
\end{table}

\begin{table}[ht!]
\begin{center}
 \begin{tabular}{ccccccccc}
	 & \multicolumn{2}{c}{$L/\lambda_s=4,L/\lambda_l=32$} 
	 & \multicolumn{2}{c}{$L/\lambda_s=8,L/\lambda_l=128$} 
	 & \multicolumn{2}{c}{$L/\lambda_s=16,L/\lambda_l=512$} 
	 & \multicolumn{2}{c}{$L/\lambda_s=4,L/\lambda_l=512$}\\
	 & \multicolumn{2}{c}{$\alpha = 5.51 \pm 0.26$} 
	 & \multicolumn{2}{c}{$\alpha = 11.69 \pm 0.32$} 
	 & \multicolumn{2}{c}{$\alpha = 28.67 \pm 0.55$} 
	 & \multicolumn{2}{c}{$\alpha = 214.16 \pm 10.83$}\\[3pt]
$p'$    & $A'$  & std$(A')$     & $A'$  & std$(A')$     & $A'$  & std$(A')$     & $A'$  & std$(A')$\\[2pt]\hline\\[-8pt]
0.001	 & 0.250 & 0.029	 & 0.228 & 0.006	 & 0.221 & 0.004	 & 0.207 & 0.005\\
0.004	 & 0.933 & 0.068	 & 0.902 & 0.015	 & 0.870 & 0.006	 & 0.817 & 0.015\\
0.007	 & 1.615 & 0.103	 & 1.562 & 0.020	 & 1.512 & 0.012	 & 1.427 & 0.020\\
0.010	 & 2.295 & 0.104	 & 2.209 & 0.029	 & 2.142 & 0.014	 & 2.029 & 0.023\\
0.013	 & 2.959 & 0.102	 & 2.852 & 0.033	 & 2.769 & 0.015	 & 2.627 & 0.024\\
0.016	 & 3.617 & 0.094	 & 3.495 & 0.036	 & 3.391 & 0.016	 & 3.220 & 0.025\\
0.019	 & 4.264 & 0.089	 & 4.133 & 0.040	 & 4.008 & 0.016	 & 3.810 & 0.029\\
0.022	 & 4.905 & 0.091	 & 4.762 & 0.044	 & 4.620 & 0.016	 & 4.396 & 0.032\\
0.025	 & 5.535 & 0.098	 & 5.386 & 0.049	 & 5.228 & 0.017	 & 4.981 & 0.033\\
0.028	 & 6.158 & 0.106	 & 6.008 & 0.056	 & 5.835 & 0.021	 & 5.563 & 0.035\\
0.031	 & 6.778 & 0.120	 & 6.626 & 0.063	 & 6.437 & 0.021	 & 6.141 & 0.036\\
0.034	 & 7.400 & 0.133	 & 7.243 & 0.068	 & 7.034 & 0.021	 & 6.717 & 0.038\\
0.037	 & 8.020 & 0.146	 & 7.856 & 0.071	 & 7.629 & 0.023	 & 7.293 & 0.039\\
0.040	 & 8.637 & 0.153	 & 8.467 & 0.072	 & 8.221 & 0.024	 & 7.864 & 0.039\\
0.043	 & 9.253 & 0.157	 & 9.072 & 0.070	 & 8.811 & 0.024	 & 8.435 & 0.040\\
0.046	 & 9.867 & 0.162	 & 9.676 & 0.071	 & 9.398 & 0.025	 & 9.005 & 0.041\\
0.049	 & 10.477 & 0.172	 & 10.277 & 0.071	 & 9.981 & 0.026	 & 9.571 & 0.039\\
0.052	 & 11.085 & 0.181	 & 10.875 & 0.072	 & 10.562 & 0.028	 & 10.137 & 0.039\\
0.055	 & 11.691 & 0.188	 & 11.470 & 0.073	 & 11.137 & 0.031	 & 10.701 & 0.040\\
0.058	 & 12.293 & 0.190	 & 12.061 & 0.075	 & 11.713 & 0.032	 & 11.264 & 0.041\\
 \end{tabular}
 \end{center}
 \caption{\label{tab:4}Sampled numerical results for the contact area fraction $A'$ growth with the normalized nominal pressure $p' = p_0/(\sqrt{2m_2}E^*)$ for surfaces with $H=0.8$ and different cutoffs in the spectrum, mean results over 10 realizations is shown as well as the standard deviation denoted std$(A')$. The contact area is corrected using~\eqref{eq:cor_area}.}
\end{table}

\begin{figure}[htb!]
 \includegraphics[width=1\textwidth]{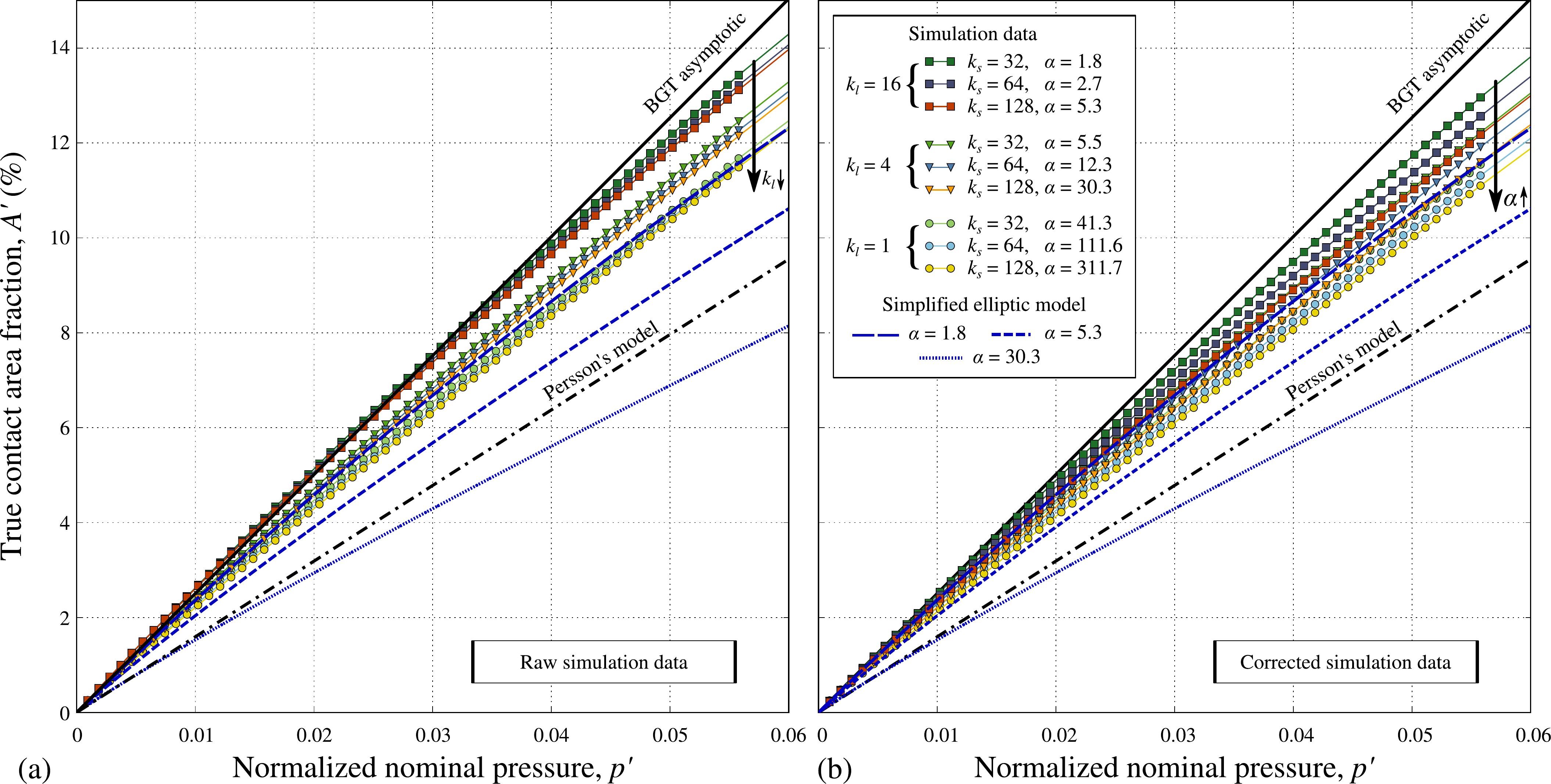}
 \caption{\label{fig:area_ijss}Post-treated data from~\cite{yastrebov2015ijss}. True contact area evolution with the normalized nominal pressure $A'(p')$: 
 (a) raw simulation data as presented in~\cite{yastrebov2015ijss}, (b) corrected simulation data obtained using~\eqref{eq:cor_area}.
 Simulation results (lines with points) are compared with analytic models: Persson's model (dash-dotted line), asymptotic linear solution of asperity models, denoted BGT asymptotic~\cite{bush1975w,carbone2008jmps} (solid line) and Greenwood's simplified elliptic model~\cite{greenwood2006w} computed for $\alpha = 1.8$, $5.3$ and $30.3$ (dark blue dashed lines). Every numerical point corresponds to the mean value averaged over 50 simulations carried out with different rough surfaces.
 }
\end{figure}

\begin{figure}[htb!]
 \includegraphics[width=1\textwidth]{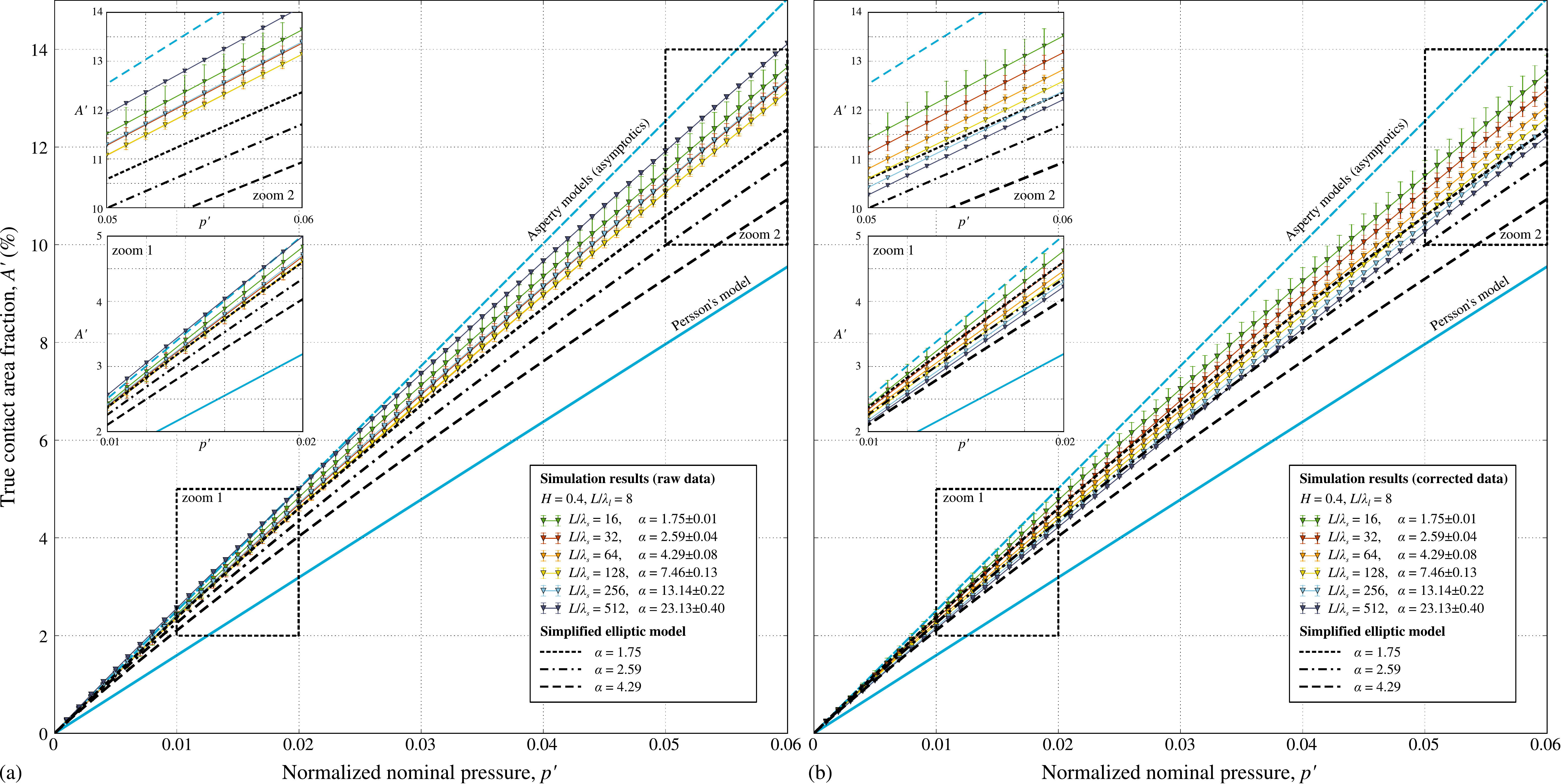}
 \caption{\label{fig:area04}True contact area evolution with the normalized nominal pressure $A'(p')$ computed for $H=0.4$: (a) raw simulation data, (b) corrected simulation data obtained using~\eqref{eq:cor_area}.
 Simulation results (lines with points) are compared with analytic models: Persson's model (solid light line), asymptotic linear solution of asperity models~\cite{bush1975w,carbone2008jmps} (dashed light line) and the Greenwood's simplified elliptic model~\cite{greenwood2006w} integrated for $\alpha = 1.75,\;2.59,\;4.29$ (black dashed and dotted lines). Two zoomed regions are shown in the insets.}
\end{figure}

\begin{figure}[htb!]
 \includegraphics[width=1\textwidth]{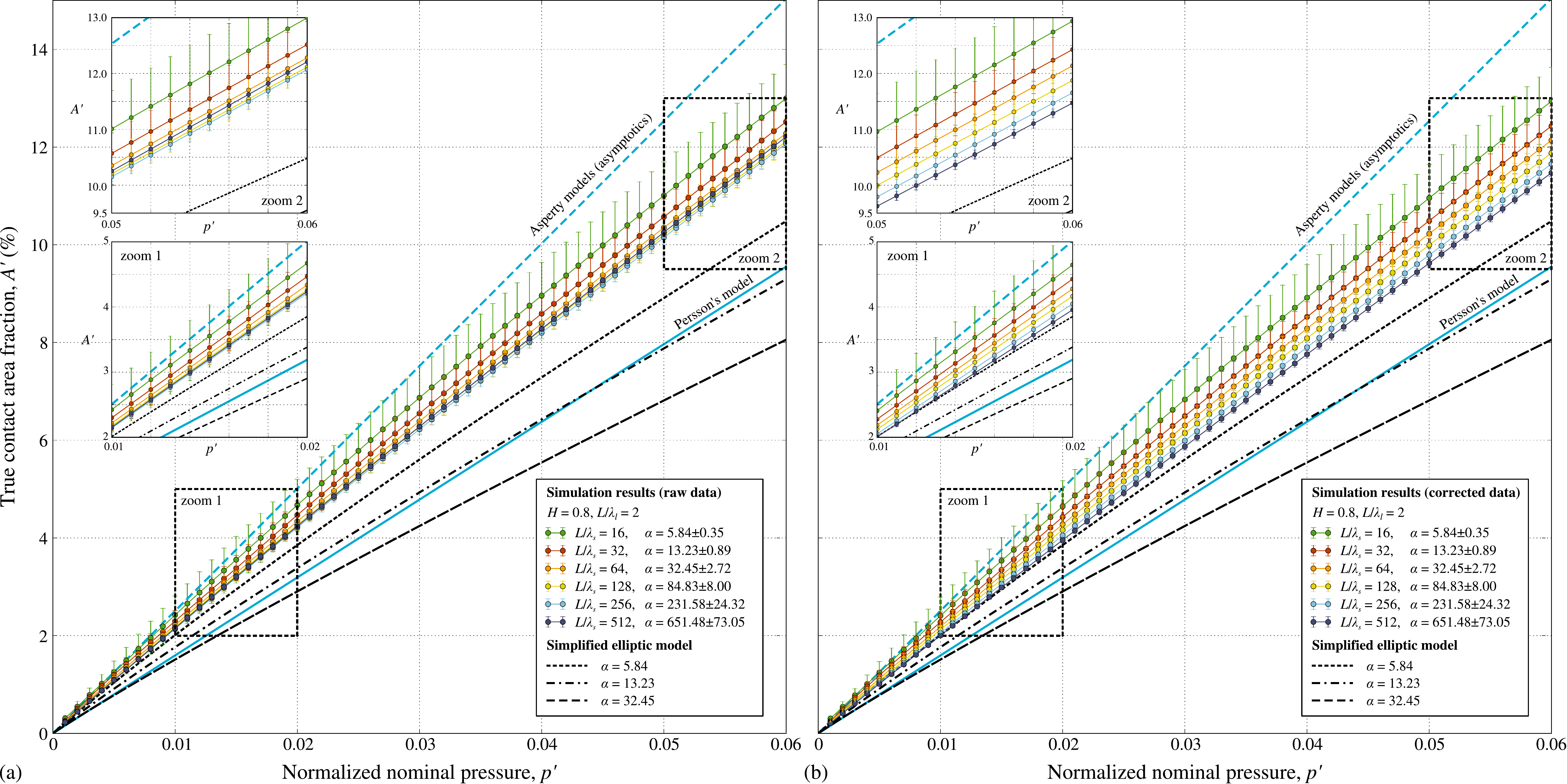}
 \caption{\label{fig:area08}True contact area evolution with the normalized nominal pressure $A'(p')$ computed for $H=0.8$: (a) raw simulation data, (b) corrected simulation data obtained using~\eqref{eq:cor_area}.
 Simulation results (lines with points) are compared with analytic models: Persson's model (solid light line), asymptotic linear solution of asperity models~\cite{bush1975w,carbone2008jmps} (dashed light line) and the Greenwood's simplified elliptic model~\cite{greenwood2006w} integrated for $\alpha = 5.84,\;13.23,\;32.45$ (black dashed and dotted lines). Two zoomed regions are shown in the insets.}
\end{figure}

In Fig.~\ref{fig:area_ijss}, the results from~\cite{yastrebov2015ijss} are post-treated using Eq.~\eqref{eq:cor_area} to obtain accurate results on the contact area evolution, both raw and corrected results are shown. The latter demonstrate a reasonable dependence on the Nayak parameter, the spurious dependence on the cutoff wavenumbers, present in the raw data, is corrected. In Figs.~\ref{fig:area04},\ref{fig:area08}, some additional results on the contact area evolution are presented for $H=0.4$ and $H=0.8$, respectively. Again, both the raw and corrected data are presented, demonstrating the correcting effect of Eq.~\eqref{eq:cor_area} suggested in~\cite{yastrebov2016w}.

%
%

\newpage


\end{document}